\documentclass[a4paper,11pt]{article}
\pdfoutput=1
\usepackage{graphicx}
\usepackage{epsf,amsmath,bbold,amsfonts,stmaryrd}

\usepackage[utf8]{inputenc}
\usepackage{mathrsfs}
\usepackage{appendix}
\usepackage{amssymb}
\usepackage{float}
\usepackage{color}
\usepackage{cite}
\usepackage{hyperref}
\hypersetup{pageanchor=false}
\usepackage{indentfirst}
\usepackage{url}
\usepackage{float}
\usepackage{caption}
\usepackage[numbers,square,comma,sort&compress,merge]{natbib}

\def\a{\alpha}

\def\l{\lambda}

\def\mc{\mathcal}

\def\nn{\nonumber}

\def\om{\omega}
\def\Om{\Omega}

\def\p{\partial}

\def\s{\sigma}

\def\hs{\hspace}

\def\be{\begin{equation}}
\def\ee{\end{equation}}

\def\bea{\begin{eqnarray}}
\def\eea{\end{eqnarray}}

\def\ba{\begin{array}}
\def\ea{\end{array}}

\def\bc{\begin{center}}
\def\ec{\end{center}}

\def\bl{\begin{flushleft}}
\def\el{\end{flushleft}}

\def\br{\begin{flushright}}
\def\er{\end{flushright}}

\def\bi{\begin{itemize}}
\def\ei{\end{itemize}}

\def\bt{\begin{tabular}}
\def\et{\end{tabular}}
\newcommand{\sR}{\mathsf{R}}
\newcommand{\bb}{\mathbf { b}}
\newcommand{\rv}{\mathrm{v}}
\newcommand{\bRe}{\mathbf {R_e}}
\newcommand{\bM}{\mathbf { M}}

\numberwithin{equation}{section}

\begin{document}

\title{\textbf{Holographic Turbulence in Einstein-Gauss-Bonnet Gravity at Large $D$}}

\author{Bin Chen$^{1,2,3}$, Peng-Cheng Li$^{1}$, Yu Tian$^{4,5,6}$ and Cheng-Yong
Zhang$^{3}$\thanks{bchen01@pku.edu.cn, wlpch@pku.edu.cn, ytian@ucas.ac.cn, zhangcy0710@pku.edu.cn,
}}

\date{}

\maketitle

\vspace{-10mm}

\begin{center}
{\it
$^1$Department of Physics and State Key Laboratory of Nuclear Physics and Technology,\\Peking University, 5 Yiheyuan Road, Beijing 100871, China\\\vspace{1mm}

$^2$Collaborative Innovation Center of Quantum Matter, 5 Yiheyuan Road, Beijing 100871, China\\\vspace{1mm}

$^3$Center for High Energy Physics, Peking University, 5 Yiheyuan Road, Beijing 100871, China\\\vspace{1mm}
$^4$School of Physics, University of Chinese Academy of Sciences, Beijing 100049, China\\\vspace{1mm}
$^5$Institute of Theoretical Physics, Chinese Academy of Sciences, Beijing 100190, China\\\vspace{1mm}
$^6$  Center for Gravitation and Cosmology, College of Physical Science and Technology, Yangzhou University, Yangzhou 225009, China
}
\end{center}

\vspace{8mm}

\begin{abstract}
We study the holographic hydrodynamics in the Einstein-Gauss-Bonnet(EGB) gravity in the framework of the large $D$ expansion. We find that the
large $D$ EGB equations can be interpreted as the hydrodynamic equations describing the  conformal fluid.  These fluid equations are  truncated at the second order of the derivative expansion, similar to the Einstein gravity at large $D$. From the analysis of the fluid flows, we find that
the fluid equations can  be taken as a variant of the compressible version of the non-relativistic Navier-Stokes equations. Particularly,  in the limit of
small Mach number, these equations could be cast into the form of the incompressible Navier-Stokes equations with redefined  Reynolds number and  Mach number. By using numerical simulation, we find that  the EGB holographic turbulence shares  similar qualitative feature as the turbulence from the Einstein gravity, despite the presence of two extra terms in the equations of motion. We analyze the effect of the GB term on the holographic turbulence in detail.

\end{abstract}
\baselineskip 18pt

\thispagestyle{empty}
\newpage

\section{Introduction}
In recent years, it has been found that black hole physics simplify significantly in the limit that the number of spacetime dimensions $D$ is very large  \cite{Emparan1302, Emparan1402, Emparan1406, Emparan1502}. The key feature in the large $D$ limit  is that the gravitational field of a black hole is strongly localized near its horizon
 due to the very large radial gradient of the gravitational potential. By performing the $1/D$ expansion in the near region of the black hole, the Einstein equations can be reduced to several effective equations which capture the dynamics of the black hole. As a consequence one can solve these  equations to construct various black hole solutions, including  black string and black ring with different asymptotic behaviors and topologies. Furthermore one can study the linear stability of these solutions perturbatively to obtain the quasinormal modes, and even  study the non-linear evolutions of the solutions numerically \cite{Emparan1504, Suzuki1505,  Suzuki1506, Emparan1506, Tanabe1510,Chen1702, Tanabe1511, Rozali1607,  Miyamoto1705,Emparan:2018bmi}.

  The fact that the dynamics of the black holes are captured by the effective equations of the horizons is reminiscent of the membrane paradigm\cite{Bhattacharyya1504, Bhattacharyya1511, Dandekar1607, Dandekar1609,Bhattacharyya1704,Dandekar:2017aiv}, initialized in \cite{Damour:1978cg}\footnote{For the subsequent developments, see \cite{Thorne:1986iy}.}. One essential feature of the membrane paradigm is that the evolution of the horizon is like a viscous fluid.  In \cite{Emparan1602} it was found that the large $D$ effective equations for (AdS or asymptotically flat) black branes can be interpreted as the equations for dynamical fluid. For AdS black branes the transport coefficients were found to match well with the result obtained from the holographic study. More importantly, the study in \cite{Emparan1602} indicates that the hydrodynamical gradient expansion is truncated  at a finite order and the higher order transport coefficients are vanishing in the $D \to \infty$ limit.

 On the other hand, according to the AdS/CFT correspondence, the geometry of dynamical black holes in asymptotically anti-de Sitter spaces (AAdS)  can be described in terms of
a conformal fluid living on the conformal boundary of the spaces \cite{Bhattacharyya0712}. This fluid/gravity duality not only provides a geometric way to study fluid dynamics
 but also help to reveal new phenomena that never occurs in gravity. For instance, it is well-known that the turbulence is a ubiquitous property of the fluid if the Reynolds
 number is sufficient large. The presence of the turbulence in hydrodynamics indicates a similar turbulent behavior should appear for the AdS black holes. As expected, it was found that by numerical simulation a perturbed black brane  in AAdS
can exhibit turbulent behavior when the Reynolds number of the fluid counterparts is sufficiently large \cite{Adams1307, Green1309, Carrasco1210}.
Within the fluid/gravity duality it is possible to use the geometric tool to study  the turbulence, such as the geometric interpretation of
turbulence \cite{Adams1307}.
Therefore, the natural questions are  what is the relation between the large $D$ effective equations for the AdS black branes and the large
$D$ limit of the conformal fluid living on the AdS boundary, and can the fluid equations exhibit turbulent behavior in a certain regime?

Indeed, as demonstrated by a recent work \cite{Rozali1707}, the large $D$ effective equations for the AdS black branes are equivalent to the large $D$ limit of
relativistic hydrodynamics which is naturally truncated at the second order in derivatives. Actually, in deriving the large $D$ limit of the relativistic hydrodynamics, one need to decompose $D=n+q+1$ and consider the dynamical variables  depending only on $q+1$ coordinates. Moreover one has to rescale the time and  space coordinates appropriately such that the hydrodynamics on $q+1$ dimensional system is simplified significantly in the large $n$ limit.  Such scaling laws for the coordinates  are precisely the ones used in studying the large $D$ limit of the  gravity. More interestingly the effective equations for the black brane in the large $D$ limit are exactly the same as the hydrodynamic equations of motions in a fluid frame. These equations are a variant of the compressible version of the non-relativistic Navier-Stokes equations.    Then by using analytic and numerical techniques the authors in  \cite{Rozali1707} analyzed two and three-dimensional
turbulent flow of the fluid in various regime and the relation with the geometry of the black branes. In this paper we would like to extend the study to the holographic hydrodynamics in the framework  of Einstein-Gauss-Bonnet (EGB) gravity at large $D$. 

The Einstein-Gauss-Bonnet gravity provides a nice platform  in studying the holographic hydrodynamics. According to the holographic dictionary, the higher-derivative terms in gravity may come from the stringy correction or the string interaction. The Gauss-Bonnet term,  including the quadratic terms of the curvature tensors,  appears
as the leading order correction in the low energy effective action of the heterotic string theory \cite{Zwiebach1985, Boulware1985}. Another appealing feature of the EGB gravity is that its equations of motion remain  second order such that the fluctuations around the vacuum do not have ghost-like mode. The EGB gravity has been well studied in the holographic hydrodynamics, in particular on the Kovtun-Son-Starinets(KSS) viscosity bound \cite{Kovtun:2004de}
\be
\frac{\eta}{s}\geq \frac{1}{4\pi}.
\ee
 In \cite{Brigante:2007nu,Brigante:2008gz}, it was shown that the KSS bound was violated in the EGB gravity
 \be
 \frac{\eta}{s}=\frac{1}{4\pi}(1-4\l^{D=5}_{GB}),
 \ee
 where $\l_{GB}$ is the Gauss-Bonnet coefficient in five dimensions. The causality constraints require that in five dimensions
  \cite{Hofman:2008ar,Buchel0906,Hofman:2009ug}
 \be
 -\frac{7}{36}\leq \l^{D=5}_{GB}\leq \frac{9}{100}.
 \ee
 In general dimensions $D$, the viscosity satisfies
 \be\label{etaovers}
 \frac{\eta}{s}=\frac{1}{4\pi}\Big(1-\frac{2(D-1)}{(D-3)}\l^{D}_{GB}\Big),
 \ee
 and the causality constraints require that \cite{Buchel0911}
 \be
 -\frac{(3D-1)(D-3)}{4(D+1)^2} \leq \l^{D}_{GB} \leq \frac{(D-3)(D-4)(D^2-3D+8)}{4(D^2-5D+10)^2}.\label{causal}
 \ee

 The large $D$ study of the EGB gravity \cite{Chen1511, Chen1703, Chen1707} shares many similar features with that of the Einstein gravity. For example,  there are also two classes of quasinormal modes for the EGB black holes: one consists of the decoupled modes, which characterize the dynamics of the black hole,  and  the other one consists  of featureless non-decoupling modes \cite{Chen1511}. For the EGB black strings (branes), the final states of the non-linear evolution  are non-uniform black strings if the strings are thin enough. This is qualitatively the same as the behavior of black strings in the Einstein gravity.
Besides, the large $D$ effective equations for the (asymptotically flat) EGB black branes can
be interpreted as the fluid equations as well\cite{Chen1707}. So for the AdS black branes in the EGB gravity one may  expect that the large $D$ effective equations are also of the equations
for a dynamical fluid, which allows us to study the holographic turbulence
in the EGB gravity.

In the paper, based the work of \cite{Rozali1707} we will study the  holographic turbulence in the EGB gravity by using the large $D$ expansion method. Unlike the case in the Einstein gravity, up to second order the hydrodynamics dual to the EGB gravity is unknown. In this work, we focus on the holographic EGB hydrodynamics.
As we show in section \ref{section:equationofmotion}, if we take the same scalings for the coordinates as those used in  constructing the EGB black strings \cite{Chen1707}  at large $D$, then we obtain the effective equations which
have the form of the hydrodynamical equations naturally truncated at second order in derivatives. Moreover, the transport coefficients are the same as those obtained from AdS/CFT \cite{Buchel0911}.

In section \ref{section:analyticalanalysis}  we give an analytical discussion on the  the large $D$ EGB fluid flows. We show that in the small Mach number limit, the EGB equations could be reduced to the incompressible Navier-Stokes equations with modified Reynolds and Mach number. However, if the Mach number is not small, the extra terms coming from the GB term may play an important role, especially in 2D flow.
Then in section \ref{section:numericalstudy} we numerically solve the equations of motion  and analyze the effect of the GB term on the turbulence. We surprisingly find that the extra terms play negligible role in the evolution of the turbulence. We find that the 2D EGB turbulence has qualitatively similar behavior as the one from the Einstein gravity. For example, we notice an inverse and direct cascade in the turbulence, with the energy spectrum of the direct cascade obeying a $k^{-4}$ power law.
 In section \ref{section:geometricinter} we verify that the relation between the horizon
 curvature power spectrum and the hydrodynamic energy power spectrum proposed by \cite{Adams1307} holds for the EGB gravity. We end with a summary and some discussions in section \ref{section:summary}.

\section{Equations of motion of relativistic hydrodynamics}\label{section:equationofmotion}

\subsection{$1/n$ expansion of the EGB equations }
The action of the $D$ dimensional EGB gravity with a negative cosmological constant $\Lambda=-(D-1)(D-2)/2$ (we have set the radius of AdS space to unity) is given by
\be
I=\frac{1}{16\pi G}\int d^Dx\sqrt{-g}\biggl(R-2\Lambda+\alpha L_{GB}\biggl),
\ee
with
\be
L_{GB}=R_{\mu\nu\lambda\delta}R^{\mu\nu\lambda\delta}-4R_{\mu\nu}R^{\mu\nu}+R^2,
\ee
where $\alpha$ is the GB coefficient. Here we follow the conventions in \cite{Chen1511, Chen1703, Chen1707}. The GB coefficient $\a$ is related to the one used in \cite{Buchel0911} by
\be
\a=\frac{1}{(D-3)(D-4)}\l^D_{GB}.
\ee
As the GB term appear in the low energy effective action   in the heterotic string theory \cite{Boulware1985}, the GB coefficient is usually taken to be positive. However,  just taking the EGB theory as a gravity theory, there is no reason to restrict the GB coefficient to be positive. Actually from the causal constraints \cite{Buchel0911} it can be negative as well. In this work, we do not make any constraints on the GB coefficient.

 From the action, we obtain the equations of motion for the metric
\be\label{EGBequations}
R_{\mu\nu}-\frac{1}{2}g_{\mu\nu}R+\Lambda g_{\mu\nu}+\alpha H_{\mu\nu}=0,
\ee
where
\be
H_{\mu\nu}=-\frac{1}{2}g_{\mu\nu}L_{GB}+2(RR_{\mu\nu}-2R_{\mu\gamma}R^{\gamma}_{\,\,\nu}+2R^{\gamma\delta}R_{\gamma\mu\nu\delta}
+R_{\mu\gamma\delta\lambda}R_{\nu}^{\,\,\gamma\delta\lambda}).
\ee
The black brane solution in EGB gravity can be written as \cite{Cai0109}
\be
ds^2=-r^2 h(r)dt^2+\frac{dr^2}{r^2h(r)}+\frac{r^2}{L_c^2}(\delta_{ij}dx^idx^j),\quad i,j=1,\cdots, D-2,
\ee
where
\bea
h(r)&=&\frac{1}{2\tilde{\alpha}}\Bigg(1-\sqrt{1+\frac{4\tilde{\alpha}r_+^{D-1}}{r^{D-1}}-4\tilde{\alpha}}\,\Bigg),\label{blackbrane}\\
\tilde{\alpha}&=&\alpha(D-3)(D-4)=\l^D_{GB},\label{blackbrane:alpha}
\eea
 $r_+$ is the horizon radius, and $\tilde{\alpha}$ is exactly the Gauss-Bonnet coefficient used in \cite{Buchel0911}. Note that in the above expression  $L_c$ is introduced as the effective radius of the AdS space in EGB gravity, since as $r\to \infty$, $h(r)\to1/L_c^2$, with
\be
L_c^2=\frac{1+\sqrt{1-4\tilde{\alpha}}}{2}.
\ee
From above we can find that there exists an upper bound for the GB coefficient $\tilde{\alpha}$, i.e.
\be\label{upperboundforalpha}
\tilde{\alpha}\leq\frac{1}{4},
\ee
beyond which the theory is not well-defined. On the other hand, the causality constraint (\ref{causal}) requires that in the large D limit,
\be
 -\frac{3}{4}\leq \tilde{\alpha} \leq 1/4.
 \ee
It turns out that the upper bound is the same as (\ref{upperboundforalpha}).

Using the Eddington-Finkelstin coordinates the solution (\ref{blackbrane}) can be written as
\be
ds^2=-r^2 h(r)dv^2+2dvdr+\frac{r^2}{L_c^2}(\delta_{ij}dx^idx^j).
\ee
 $L_c$ can be absorbed into the  redefinition of $x^i$, i.e. $x^i\to x^i/L_c$, such that $L_c$ does not appear in the metric explicitly.
This form of the metric leads to   the following metric ansatz
\bea\label{metricansatz}
ds^2&=&-r^2A(v,r,z^c) dv^2-2u_v(v,r,z^c) dvdr-2u_a(v,r,z^c) dz^a dr\nonumber\\
&&+r^2\Big(-2C_a(v,r,z^c) dz^a dv+G_{ab}(v,r,z^c)dz^adz^b +d\vec{x}^2\Big),
\eea
where $a, b, c=1, \cdots q$, $\vec{x}$ is an $n$ dimensional vector and $D=n+q+2$. Note that as in \cite{Emparan1506, Rozali1707}  we keep $q$ finite and let $n\to\infty$. In this
case most of the spatial directions remain translational invariance, so that we only study the deformations of the black brane that depend  only on a finite number of coordinates $z^a$. Moreover, in the following we use $1/n$ rather than $1/D$ as the expansion parameter.

In order to perform the $1/n$ expansion properly we need to specify the large $D$ scalings of the metric functions. According to the properties of
the boundary hydrodynamics \cite{Buchel0911}, the speed of sound of long-wavelength perturbations is of $\mc O(1/\sqrt{n})$, this indicates that to capture the physics of
the long-wavelength  perturbations, we should rescale
$z^a\to z^a/\sqrt{n}$, $\vec{x}\to\vec{x}/\sqrt{n}$.
In addition, we consider small velocities $\mc O(1/\sqrt{n})$ along the brane directions. Therefore, the large $n$ scalings of the metric functions are respectively
\be
A=\mc O(1), \quad u_v=\mc O(1),\quad u_a=\mc O(n^{-1}),\quad C_a=\mc O(n^{-1}),\quad G_{ab}=\frac{1}{n}\Big(\delta_{ab}+\mc O(n^{-1})\Big).
\ee
These scalings are exactly the same as the one in discussing the black string in the large $D$ EGB theory\cite{Chen1707}.
We can rewrite the metric ansatz (\ref{metricansatz}) as
\be
ds^2=-r^2A dv^2-2u_v(v,r,z^c) dvdr+r^2\Big(-\frac{2}{n}\tilde{C}_a dz^a dv+\frac{1}{n}\tilde{G}_{ab}dz^adz^b +\frac{1}{n}d\vec{x}^2\Big),
\ee
where by gauge choice we have set $u_a=0$ and all the functions in the above expression are of $\mc O(1)$ now.

At large $D$ the radial gradient becomes dominant, i.e. $\partial_r=\mc O(n)$, $\partial_v=\mc O(1)$, $\partial_a=\mc O(1)$, so in the near region of the black brane
it is  better to introduce a new radial coordinate
\be
\sR=\Big(\frac{r}{r_0}\Big)^n,
\ee
such that the derivative with respect to $\sR$ is finite in the large $D$ limit, where $r_0$ is a horizon length scale which can be set to be unity $r_0=1$.
To solve the EGB equations we need to specify the boundary conditions at large $\sR$, which are
\be
A=L_c^{-2}+\mc O(\sR^{-1}),\quad \tilde{C}_a=\mc O(\sR^{-1}),\quad \tilde{G}_{ab}=\delta_{ab}+\mc O(\sR^{-1}).
\ee
On the other hand, the solutions have to be regular at the horizon.


At the leading order of the $1/n$ expansion, the EGB equations only contain $\sR$-derivatives so they can be solved by performing $\sR$-integrations. After imposing the boundary conditions the leading order solutions are obtained as
\be
A=\frac{1}{2\tilde{\alpha}}\Big(1-\bb\,\Big),\quad u_v=-1,
\ee
\be
\tilde{C}_a=\frac{p_a(v,z^a)\beta}{2\tilde{\alpha}\, m(v,z^a)}\Big(\bb-\beta\Big),\quad\tilde{G}_{ab}=\delta_{ab}+\frac{G_0}{n},
\ee
where
\bea\label{G0}
G_{0}&=&\frac{2\beta}{(1+\beta)}\Biggl[\frac{\pi}{4}-\arctan\frac{\bb}{\beta}
+\ln\frac{\bb+\beta}{2\beta}
-\frac{\beta}{2}\ln\frac{\bb^2+\beta^2}{2\beta^2}\Biggl]\times\nonumber\\
&&\frac{\Big(p_a\partial_b m+p_b\partial_a m-m(\partial_b p_a+\partial_a m_b)\Big)}{m^2}+\frac{(\bb-\beta)\beta^2p_a p_b}{(1+\beta) \tilde{\alpha} m^2},
\eea
and to simplify the expressions we have introduced the quantities $\bb$ and $\beta$
\bea\label{bandbeta}
\bb&=&\sqrt{1+\frac{4\tilde{\alpha} m(v,z^a)}{\sR}-4\tilde{\alpha}},\\
\beta&=&\sqrt{1-4\tilde{\alpha}},\quad\mathrm{with} \quad 0\leq\beta\leq2.
\eea
As shown in \cite{Emparan1602} and \cite{Chen1707}, the $1/n$ terms in $\tilde{G}_{ab}$ are obtained at the next-to-leading order in the $1/n$ expansion
of the EGB equations. It must be included since it also appears in the EGB equations at the leading order of the $1/n$ expansion. $m(v,z^a)$ and $p_a(v,z^a)$
 are  the integration functions of $\sR$-integrations of the EGB equations.
At the next-to-leading order of the $1/n$ expansion, $m$ and $p_a$ must satisfy the effective equations
\be\label{effeq1}
\partial_v m-\partial_b\partial^b m=-\frac{2\beta}{\beta+1}\partial_b p^b,
\ee
\bea\label{effeq2}
&&\partial_v p_a-\frac{(\beta^2+1)}{(\beta+1) }\partial_b\partial^b p_a
-\frac{\beta(\beta-1)}{(\beta+1)}\partial_a\partial^b p_b+\frac{\beta(\beta-1)}{(\beta+1)m}p_a\partial_b\partial^b m
+\frac{\beta(\beta-1)}{(\beta+1)m}p_b\partial^b\partial_a m\nonumber\\
&&+\frac{2\beta}{(1+\beta)}\partial_b\Big(\frac{p^bp_a}{m}\Big) -\frac{\beta(\beta-1)}{(1+\beta)m^2}(p_b\partial^bm\partial_a m+p_a \partial^bm\partial_b m)\nonumber\\
&&+\frac{\beta(\beta-1)}{(1+\beta)m}(\partial^bm\partial_bp_a+\partial_am\partial^bp_b)+\frac{\partial_a m}{\beta}=0,
\eea
In the limit $\tilde{\alpha}\to0$, i.e. $\beta\to1$, the above equations reproduce the ones in the Einstein gravity \cite{Emparan1602}.
\subsection{Linear analysis}\label{subsection:linearanalysis}
Considering a small perturbation around  the static uniform black brane solution, i.e. $m=m_0=$ const., $p_a=0$,
\bea
m&=&m_0+\delta m e^{-i\omega v+ i k_b z^b},\\
p_a&=&\delta p_a e^{-i\omega v+ i k_b z^b},
\eea
and plugging these into the effective equations (\ref{effeq1}) and (\ref{effeq2}), we can read the quasinormal mode frequencies of the
shear mode and the sound mode.

\paragraph{Shear mode.} The frequency is
\be\label{QNMfrequency}
\omega=-i\frac{\beta^2+1}{\beta+1} k^2.
\ee
As we will see in the next subsection, this dispersion relation is related to the transport coefficient of the viscous hydrodynamics.

\paragraph{Sound modes.} The frequencies are
\be
\omega_{\pm}=-i\frac{\beta^2+1}{\beta+1} k^2\pm \frac{k\sqrt{2(\beta+1)-\beta^2(\beta-1)^2k^2}}{\beta+1}.
\ee
It is easy to see that the perturbation is always stable. Up to the leading order of $k$, one finds $\omega_{\pm}=\pm \sqrt{\frac{2}{\beta+1}}k$. As we will see in the next subsection,
this coefficient corresponds to the speed of sound of the long-wavelength perturbation of the fluid.

\subsection{Dynamical fluid}
In \cite{Emparan1602}, it was shown that the effective equations for the black branes can be interpreted as the equations for a dynamical fluid.
This turns out to be true for the asymptotically flat EGB black branes  as well\cite{Chen1707}. Here we show that  the effective equations (\ref{effeq1}) and (\ref{effeq2})
can be actually transformed into the fluid equations.
Firstly, we introduce $\rv_a$ by
\be
p_a=\frac{1+\beta}{2\beta}\Big(m \rv_a+\partial_a m\Big),
\ee
then we find that the first effective equation (\ref{effeq1}) becomes the continuity equation for the energy density
\be\label{fluideq1}
\partial_v m+\partial^a( m \rv_a)=0,
\ee
with $m$ and $\rv_a$ being taken as the energy density and the velocity of the fluid flows.

Furthermore,  we notice that (\ref{effeq2}) can be written in terms of $\rv_a$ as

\be\label{fluideq2}
\partial_v(m \rv^a)+\partial_b(m \rv^a \rv^b+\tau^{ab})=0
\ee
where
\be\label{stresstensor}
\tau_{ab}=\frac{2}{\beta+1}\delta_{ab} m-
\frac{2(\beta^2+1)}{\beta+1} m\partial_{(a} \rv_{b)}-\frac{2\beta^2-\beta+1}{\beta+1}m\partial_a\partial_b\ln m.
\ee
This equation is for the momentum conservation, with $\tau_{ab}$ being the stress tensor. Therefore, we may  interpret the effective equations as the
equations of motion for non-relativistic\footnote{In fact, since the physical velocity for the fluid flow is $\mathbf{v}^i=\mathrm{v}^i/\sqrt{n}$,
 the fluid flow is non-relativistic  in the large $n$ limit. In \cite{Rozali1707}, due to the extra factor $1/\sqrt{n}$ carried by the velocity filed,
  the large $D$ hydrodynamics is shown to be non-relativistic.}, compressible fluid naturally truncated at second order in derivatives.

As shown in \cite{Rozali1707},  the Einstein equations at large $D$ are equivalent to the large $D$ hydrodynamics within the fluid/gravity duality.  This
statement holds for the EGB gravity as well:  the EGB equations
at large $D$ are equivalent to the large $D$ hydrodynamics describing the conformal fluid living on the conformal boundary of the AdS space. Although
the relativistic hydrodynamics that dual to AdS black branes in the EGB gravity up to second order is unknown at present, we can still find some hints by comparing the
properties of the fluid flows (\ref{fluideq1}) and (\ref{fluideq2}) with the results obtained from AdS/CFT \cite{Buchel0911}.
From (\ref{stresstensor}) we find the pressure
\be
P=\frac{2}{\beta+1}m,
\ee
which gives the equation of state of the black string. The speed of sound of the long-wavelength perturbations is then
\be\label{soundspeed}
c_s=\sqrt{\frac{\partial P}{\partial m}}=\sqrt{\frac{2}{\beta+1}}.
\ee
Due to the scaling relation we used in the metric ansatz, the physical speed of sound is $\mathbf{c_s}=c_s/\sqrt{n}$,  which
 is small in the large $n$ limit.
Moreover, the shear viscosity is
\be\label{viscosity}
\eta=\frac{\beta^2+1}{\beta+1} m,
\ee
then the  ratio of the shear viscosity to the entropy density is given by
\be
\frac{\eta}{s}=\frac{1}{4\pi}\frac{\beta^2+1}{\beta+1},
\ee
with the entropy density $s=4\pi m$. Both this result and the speed of sound are in accord with the results found previously in \cite{Buchel0911} by taking the large $D$ limit
and using our convention\footnote{In fact, if we use the convention in \cite{Buchel0911} and for simplicity only consider one momentum along the brane direction, then we find the effective equations
\be
\partial_v m-\frac{1+\beta}{2}\partial_z^2 m+\beta \partial_z p_z=0,
\ee
\be
\partial_v p_z+\frac{-1+\beta-2\beta^2}{2}\partial_z^2 p_z+\frac{\beta(\beta-1)p_z}{m}\partial_z^2 m
+\frac{\beta(\beta-1)}{m}\partial_z p_z \partial_z m
+\Big(\frac{1}{\beta}-\frac{\beta p_z^2}{m^2}\Big)\partial_z m+\frac{2\beta p_z\partial_z p}{m}=0.
\ee
In terms of  $p_z=\frac{1}{\beta}m v_z+\frac{1+\beta}{2\beta}\partial_z m$,  the effective equations have the similar form as
(\ref{fluideq1}) and (\ref{fluideq2}). In this case  the ratio of the shear viscosity to the entropy density is found to be
$\frac{\eta}{s}=\frac{1}{4\pi}\frac{\beta^2+1}{2}$ which is exactly the same as the large $D$ limit of (\ref{etaovers}).
 }.

\section{General analysis of the large $D$ EGB fluid flows}\label{section:analyticalanalysis}

In this section we give a general discussion on the holographic EGB fluid flows.  We will  relate the equations of motion (\ref{fluideq1}) and (\ref{fluideq2})
 to the compressible Navier-Stokes equations. In terms of
\be
f^a=\frac{2\beta}{\beta+1}\frac{p^a}{m},
\ee
the equations of motion (\ref{fluideq1}) and (\ref{fluideq2}) can be rewritten as\footnote{Note that $\nabla\vec{f}\cdot\nabla m$ is different from $\nabla m\cdot\nabla \vec{f}$: the former one is written as $\partial_a f^b\partial_bm$ but the latter one is $\partial_bm \partial^bf_a$. }
\be\label{feq1}
\partial_v m+\vec{f}\cdot\nabla m+m \nabla\cdot \vec{f}=\nabla^2 m,
\ee
\bea\label{feq2}
&&\partial_{v}\vec{f}+(\vec{f}\cdot\nabla)\vec{f}-\frac{\beta^{2}+1}{\beta+1}\nabla^{2}\vec{f}-\frac{2(\beta^{2}+1)}{\beta+1}(\frac{\nabla m}{m}\cdot\nabla)\vec{f}\nonumber\\
 &&+\frac{2}{\beta+1}\frac{\nabla m}{ m}-\frac{\beta(\beta-1)}{\beta+1}\nabla(\nabla\cdot\vec{f})+\frac{\beta(\beta-1)}{\beta+1}\Big(\nabla\vec{f}\cdot\frac{\nabla m}{m}-\frac{\nabla m}{m}\cdot\nabla \vec{f}\Big)=0.
 \eea
 It is convenient to introduce the following dimensionless variables,
 \be\label{dimensionlessvars}
 \vec{u}=\frac{\vec{f}}{U},\quad \epsilon=\frac{m}{E},\quad\frac{\partial}{\partial\widetilde{v}}=\frac{L_0}{U}\frac{\partial}{\partial v},\quad \widetilde{\nabla}=L_0 \nabla,
 \ee
 where $L_0$ is a characteristic length scale of the system, $U$ is a characteristic velocity and $E$ is a characteristic
 energy density. In terms of these dimensionless variables, (\ref{feq1}) and (\ref{feq2}) become of the forms
 \be\label{Eq:dimenless1}
\partial_{\widetilde{v}} \epsilon+ \vec{u}\cdot\widetilde{\nabla} \epsilon+\epsilon \widetilde{\nabla}\cdot  \vec{u}=\frac{1}{R_e}\widetilde{\nabla}^2 \epsilon,
\ee
 \bea\label{Eq:dimenless2}
&&\partial_{\widetilde{v}}\vec{u}+(\vec{u}\cdot \widetilde{\nabla})\vec{u}-\frac{\beta^{2}+1}{\beta+1}\frac{1}{Re} \widetilde{\nabla}^{2}\vec{u}-\frac{2(\beta^{2}+1)}{\beta+1}\frac{1}{R_e} (\frac{ \widetilde{\nabla} \epsilon}{\epsilon}\cdot \widetilde{\nabla})\vec{u}\nonumber\\
 &&+\frac{2}{\beta+1}\frac{1}{M^2}\frac{ \widetilde{\nabla} \epsilon}{ \epsilon}-\frac{\beta(\beta-1)}{\beta+1}\frac{1}{R_e} \widetilde{\nabla}( \widetilde{\nabla}\cdot\vec{u})+\frac{\beta(\beta-1)}{\beta+1}\frac{1}{R_e}\Big( \widetilde{\nabla}\vec{u}\cdot\frac{ \widetilde{\nabla} \epsilon}{\epsilon}-\frac{ \widetilde{\nabla} \epsilon}{\epsilon}\cdot \widetilde{\nabla} \vec{u}\Big)=0,
 \eea
 where
\be
R_e=L_0\, U,\quad M=U,
\ee
are the Reynolds number and  the Mach number. Although the above equations are slightly more complicated than the ones  appearing in the Einstein gravity,
they can actually be taken  as a variant of the compressible version of the non-relativistic Navier-Stokes equations. This can be seen by
taking the incompressible limit and redefining the Reynolds number and the Mach number.

Let us first consider the simple case when the Mach number is small, and the fluid flow is nearly incompressible. We can expand $\vec{u}$ and $\epsilon$ in series of $M$, i.e.
 \be
 \vec{u}=\sum_{i=0}M^i \vec{u}_{(i)},\quad \epsilon=\sum_{i=0}M^i\epsilon_{(i)}.
 \ee
 Then from the above equations we find  that $\epsilon_{(0)}$ and $\epsilon_{(1)}$ are constant, and $\vec{u}_{(0)}$
 and $\epsilon_{(2)}$ satisfy the following equations,
 \be\label{incomNS1}
 \nabla\cdot \vec{u}_{(0)}=0,\quad \partial_{\widetilde{v}}\vec{u}_{(0)}+(\vec{u}_{(0)}\cdot  \widetilde{\nabla})\vec{u}_{(0)}-\frac{\beta^{2}+1}{\beta+1}\frac{1}{R_e} \widetilde{\nabla}^{2}\vec{u}_{(0)}
 +\frac{2}{\beta+1}\frac{ \widetilde{\nabla} \epsilon_{(2)}}{ \epsilon_{(0)}}=0,
 \ee
 with $\epsilon_{(0)}$ being the density and $\epsilon_{(2)}$ being the pressure.
 If we redefine the  Reynolds number and the Mach number as
 \be\label{ReandM}
\bRe=\frac{\beta+1}{\beta^{2}+1} L_0\, U,\quad \bM=\sqrt{\frac{\beta+1}{2}}U,
 \ee
 then  instead of (\ref{incomNS1}),  $\vec{u}_{(0)}$
 and $\epsilon_{(2)}$ satisfy
\be
 \nabla\cdot \vec{u}_{(0)}=0,\quad \partial_{\widetilde{v}}\vec{u}_{(0)}+(\vec{u}_{(0)}\cdot  \widetilde{\nabla})\vec{u}_{(0)}
 +\frac{ \widetilde{\nabla} \epsilon_{(2)}}{ \epsilon_{(0)}}=\frac{1}{\bRe} \widetilde{\nabla}^{2}\vec{u}_{(0)},\label{NS}
 \ee
which are exactly the same as the incompressible Navier-Stokes equations. In this case, it should be appropriate to view $\bRe$ and
$\bM$ as the Reynolds number and the Mach number of our large $D$ fluid flow. In fact, the Reynolds number and the Mach number  in the viscous fluid are defined as
\be
\bRe=\frac{m L_0 U}{\eta},\quad \bM=\frac{U}{c_s},
\ee
where $\eta$ and $c_s$ are the shear viscosity and the sound velocity. From the relations (\ref{viscosity}) and (\ref{soundspeed}), it is easy to see that
 they are exactly the same as (\ref{ReandM}).

 Up to now, we have found that in the small Mach number limit, the equations of motion for the holographic fluid flow in the EGB theory take exactly the form of the Navier-Stokes equations for an incompressible fluid. The impact of the GB term on the fluid flow is totally encoded in the redefinition of the Reynolds number and the Mach number. The incompressible flow has been well studied (see for example in the textbook \cite{Davidson2015}), and here we just give a brief review on the main results for the freely decaying fluid turbulence.

 One may define the kinetic energy and the enstrophy by
 \be
 E_I=\frac{1}{2}\int |\vec u_{(0)}|^2d^qx, \hs{3ex}\Om_I=\frac{1}{2}\int \om_{(0)ij}\om_{(0)}^{ij}d^qx,
 \ee
 where
 \be
 \om_{(0)ij}=\p_iu_{(0)j}-\p_ju_{(0)i}
 \ee
  is the vorticity 2-form. It is illuminating to write the kinetic energy and the enstrophy in terms of the energy spectrum $E(k)$,
  \bea
  E_I&=&\int_0^\infty E(k)dk \nn\\
  \Om_I&=&\int_0^\infty k^2E(k)dk,
  \eea
The energy spectrum $E(k)$ provides a measure of how the energy  is distributed across the various eddy size $\sim 1/k$.
From the equations (\ref{NS}), one can get the evolution equations for $E_I$ and $\Omega_I$
 \bea
 \partial_{\widetilde{v}}E_I&=&-\frac{1}{\bRe}\Om_I, \\
 \partial_{\widetilde{v}}\Om_I&=&\int \om^{(0)ij}\om_{(0)jk}\s_{(0)}^{ki}d^qx-\frac{1}{\bRe}P_I, \label{Omevolution}
 \eea
 where $\s_{(0)}$ and $P_I$ are the non-relativistic shear tensor and the palinstrophy, defined by
 \be
  \s_{(0)ij}=\p_iu_{(0)j}+\p_ju_{(0)i},\hs{3ex}P_I=\frac{1}{2}\int \p_k\om_{(0)ij}\p^k\om_{(0)}^{ij}d^qx
 \ee
 As $\Om_I \geq 0$, the energy $E_I$ is always decreasing. For the enstrophy equation (\ref{Omevolution}), the two terms on the right-hand side correspond to the the generation
 of the enstrophy via the vortex line stretching and the destruction of the enstrophy by the viscous forces. In the large Reynolds number limit, there is an approximate balance between the production  and the viscous dissipation of the enstrophy. Thus, to keep $\Om_I$ a constant, the energy spectrum has to be distributed in such a way that the energy will flow from the lower momentum modes to the higher momentum modes. Under Kolmogorov's Similarity Hypothesis,  the energy spectrum
 behaves as
 \be
 E(k)\sim \varepsilon^{2/3} k^{-5/3},
 \ee
 which is known as the Kolmogorov's five-thirds law, where $\varepsilon$ denotes the rate of energy dissipation at small scales.

 The turbulence in two spatial dimensions is a little bit special.
 In this case, the vortex stretching term vanishes for an incompressible flow, and the enstrophy declines monotonically as the flows evolves so that it is bounded from above by its initial value. In the large Reynolds number limit, the energy is conserved. It turns out that there could be another kind of energy cascade, the so-called inverse energy cascade, in which the energy flow to the lower momentum modes. Besides, before the turbulence adjusts to
 its fully developed state, there is a direct cascade of the enstrophy from the large scales down to the small scales due to the continual filamentation of the vorticity. In the inertial range, according to Batchelor \cite{Batchelor1969} , the energy spectrum for the direct cascade could be
 \be\label{directcascadelaw}
 E(k)\sim \delta^{2/3}k^{-3},
 \ee
 where $\delta$ is the rate of dissipation of the enstrophy. This is the two-dimensional analogue of the Kolmogorov's five-third law. However, unlike Kolmogorov's five-thirds law, the direct cascade law  (\ref{directcascadelaw}) is far from being well  accepted. In fact, the numerical simulations suggest that
 $E\sim k^{-n}$, where $n$ is typically a little larger than 3. For more details, one can see \cite{Davidson2015} and the references therein.
On the other hand, according to Kraichnan \cite{Kraichnan1967}, the possibility of an inverse cascade depends on the relative strengths of the non-linear
interactions between scales. If  the input of the energy is not sustained for sufficiently
long time, the inverse cascade cannot develop and there appears
 only the direct cascade in the flow. Thus, unlike the case of the forced two-dimensional flows, the inverse cascade with a  Kolomogrov-like energy scaling $\sim k^{-5/3}$ is puzzling in the
decaying case \cite{Mininni2013}.

It is illuminating to study the equations (\ref{Eq:dimenless1}) and (\ref{Eq:dimenless2}) beyond the small Mach number limit. We may define the energy and enstrophy via
\bea
E_C&=&\int \epsilon |\vec u|^2d^qx\nn\\
\Om_C&=&\int \frac{\om_{ij}\om^{ij}}{\epsilon} d^qx.
\eea
Here and in the following we need the vorticity and the shear tensor
\be
\om_{ij}=\p_iu_j-\p_ju_i,\hs{3ex}\s_{ij}=\p_iu_j+\p_ju_i.
\ee
From the equations  (\ref{Eq:dimenless1}) and (\ref{Eq:dimenless2}), we have
\bea
\partial_{\widetilde{v}}E_C&=&\frac{1}{\bM^2}\int\epsilon(\vec \nabla\cdot \vec u)d^qx-\frac{1}{4\bRe}\left(\int \epsilon(\om_{ij}\om^{ij}+\s_{ij}\s^{ij})+\cdots\right)\\
\partial_{\widetilde{v}}\Om_C&=&\int\frac{1}{\epsilon}\om_{ij}\om^{jk}\left(\s^i_k-\delta^i_k\vec \nabla\cdot \vec u\right)d^qx+(\mbox{terms proportional to $1/{\bRe}$})\label{OmC}
\eea
There are two remarkable points on these two equations. The first is that the effective Reynolds number $\bRe$ and Mach number $\bM$ appear in the equations. This suggest that the effect of the GB term could be largely encoded by these two parameters. The second is that the first term on the right hand side in the equation (\ref{OmC}) could be taken as a vortex stretching term, by which the energy is passed down the cascade to the small scales. The other terms are proportional to $1/\bRe$, and play the similar role as the
palinstrophy that characterizes the viscous dissipation at small scales.

If the Mach number is not small, the equations (\ref{Eq:dimenless1}) and (\ref{Eq:dimenless2}) are more complicated  than the ones appearing in the Einstein gravity. Actually there are two more terms proportional to $(\beta-1)$ in (\ref{Eq:dimenless2}), whose physical implication is not clear. However, both terms are inversely proportional to $1/\bRe$, and consequently  when the Reynolds number is very large, they may not play significant role on the dynamics of the flow except the effect on the viscous dissipation. In three spatial dimensions, if the Reynolds number $\bRe$ is large, the Kolmogorov cascade should appear in the turbulence as the vortex stretching term dominates. In two spatial dimensions, the vortex stretching term is vanishing. Therefore one may expect the inverse energy cascade in this case. However, due to the presence of two extra terms induced by the GB term, it is not clear if the inverse energy cascade do appear or present different feature. In order to understand the effects of the GB term, we need to do numerical simulation.


\section{Numerical Study of Turbulent Flows}\label{section:numericalstudy}

In this section  by numerically solving the equations of motion (\ref{effeq1}) and (\ref{effeq2}) we will give a  detailed analysis for  the turbulent flows in the EGB theory. In
particular, we focus on the distinctions between the Einstein gravity and the EGB theory.

As in \cite{Rozali1707}, we consider the flows in a toroidal domain of size $L$.  To solve the equations of motion (\ref{effeq1}) and (\ref{effeq2}) we use the Fourier spectral method
in the spatial directions and the fourth order Runge-Kutta method for time evolution. The initial conditions are taken  to be shear flows with small random perturbations, since in the inviscid case, i.e. $\eta\to0$, they are stationary solutions of the fluid equations. Explicitly, the initial  conditions are
\be
p_x=\delta p_x(\vec{x}),\quad p_y=\cos\Big(\frac{2\pi n_I x}{L}\Big),\quad m=m_0=\mathrm{constant},
\ee
for $q=2$, and
\bea
p_x&=&\cos\Big(\frac{2\pi n_I y}{L}\Big)+\delta p_x(\vec{x}),\quad p_y=\cos\Big(\frac{2\pi n_I z}{L}\Big)+\delta p_y(\vec{x}),\nonumber\\
p_z&=&\cos\Big(\frac{2\pi n_I x}{L}\Big)+\delta p_z(\vec{x}),\quad m=m_0=\mathrm{constant},
\eea
for $q=3$, where $\delta p_i$ denote the perturbations
\be
\delta p_i=\sum_{\vec{m}} A_{i,\vec{m}}\cos\Big(\delta\phi_{i,\vec{m}}+\frac{2\pi(\vec{m}\cdot \vec{x})}{L} \Big).
\ee
In the above expressions, $ A_{i,\vec{m}}$ denotes the amplitude of the perturbation and is chosen from a uniform distribution ranging from $0$ to a small value. The
phase $\delta\phi_{i,\vec{m}}$ is chosen from a uniform distribution ranging from $0$ to $2\pi$. $\vec{m}$ is the wavenumber in unit of $2\pi/L$. For more
details about the initial conditions, one can see \cite{Rozali1707, Green1309, Carrasco1210}.

Due to the decay caused by the shear viscosity, the shear flow we study is not steady. The Reynolds number, as a useful dimensionless quantity to predict the stability of the steady flow, may still be useful in our case. Thus the stability of the flow depends only on the instantaneous
 value of the Reynolds number \cite{ Green1309}.

For our flow, we follow the definition in \cite{Rozali1707} where the characteristic velocity appearing in (\ref{ReandM}) is given by\footnote{Note that this definition is slightly different from the one given in \cite{Green1309}, however it is easy to see that for $q=2$, they are pretty close.}
\be\label{characteristicvelocity}
U=\mathrm{max}(f)-\langle f\rangle.
\ee
$U$ characterizes the velocity fluctuation.
For the initial conditions of the flow, it is easy to obtain the Reynolds number and the Mach number
\be\label{initialRe}
\bRe_0=\frac{2\beta}{\beta^2+1}\frac{L}{m_0 n_I},\quad \bM_0=\sqrt{\frac{2\beta^2}{\beta+1}}\frac{1}{m_0},
\ee
where we have used $L_0=\frac{L}{n_I}$. From the above expressions we can see the effect of the GB term on the initial conditions. On the one hand, the presence of the
 GB term always lowers the initial Reynolds number. On the other hand, for a positive GB coefficient $\tilde{\alpha}$ (i.e. $0\leq\beta<1$), the presence of the
 GB term always lowers the Mach number, but for a negative $\tilde{\alpha}$ (i.e. $1<\beta\leq2$) it would always lead to a larger Mach number. Similar to the steady flow case, we expect that if the initial Reynolds number is sufficiently small the flow is stable against  small perturbations, and the final state is laminar. While for a sufficiently large initial Reynolds number, the flow is unstable and  we should expect the emergence of the turbulent flow.
Between these two extremes, for an intermediate initial  Reynolds  number,  there exists a critical Reynolds number \cite{Green1309}  beyond which the transition from  the laminar flow to the turbulent flow may occur, as we will see in the following.

\begin{figure}[hbtp]
        \centering
        \includegraphics[scale=0.23]{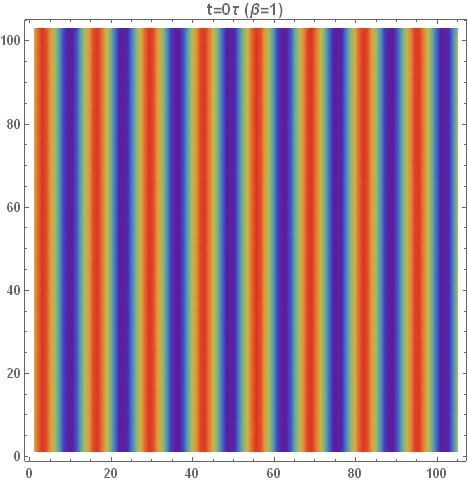}\includegraphics[scale=0.23]{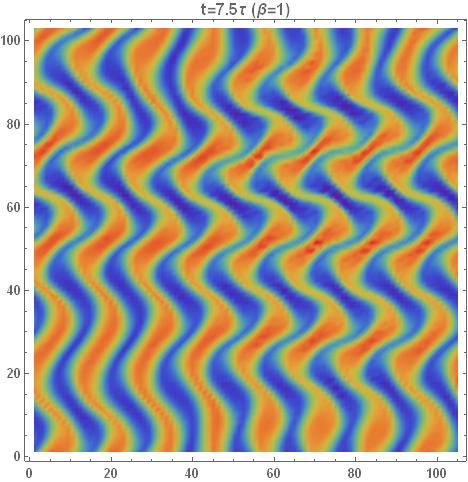}\includegraphics[scale=0.23]{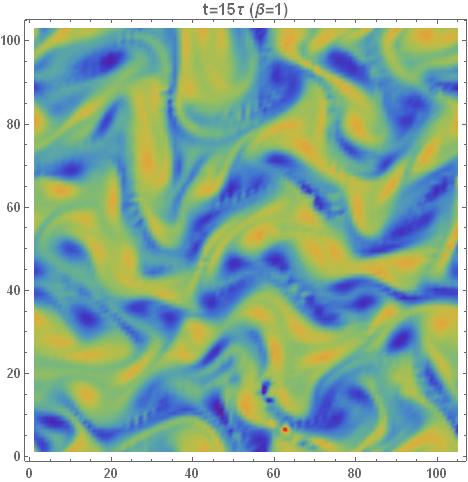}
        \includegraphics[scale=0.23]{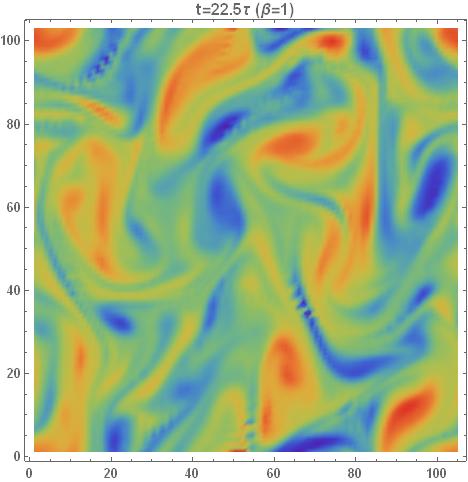}
        \includegraphics[scale=0.23]{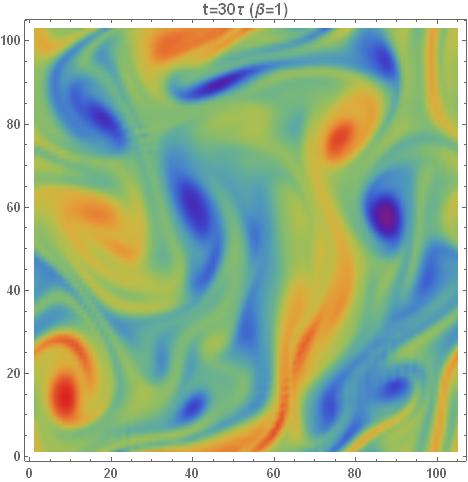}\includegraphics[scale=0.23]{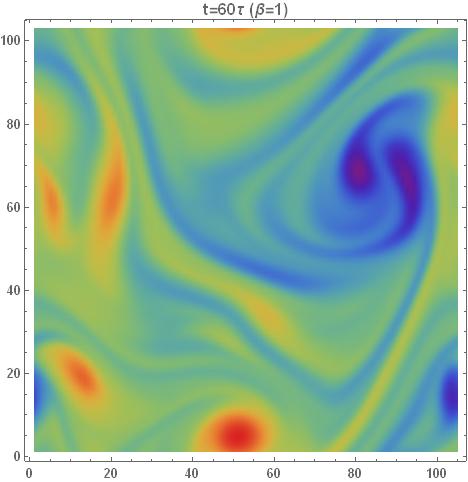}
        \includegraphics[scale=0.23]{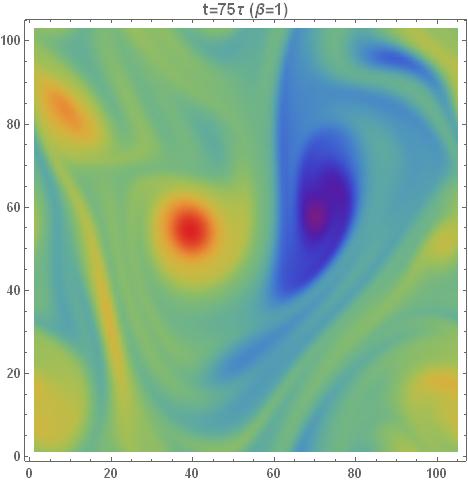}\includegraphics[scale=0.23]{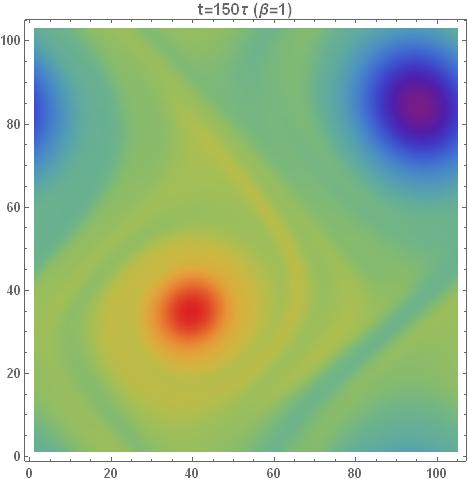}
        \caption{The vorticity field $\omega = \partial_x p_y  - \partial_y p_x$ for a two dimensional fluid flow in the Einstein gravity ($\beta=1$), with initial Reynolds number $\bRe_0 = 937.5$, $m_0=2$, initial mode $n_I=8$, and $A=10^{-5}$ at various times. We can see the initial growth of the instability, and later on  the formation of filaments which indicates the direct cascade, and then  the turbulent regime at which   a large scale structure is formed and the inverse energy cascade is apparent.   Finally, we can see the slow decay of two counter-rotating vortices. Note that here and in the following we employ $\tau$ as the units of time with $\tau=L_0 m_0$.}
        \label{fig:2Dvorticitybeta=1}
\end{figure}

\begin{figure}[hbtp]
        \centering
        \includegraphics[scale=0.2325]{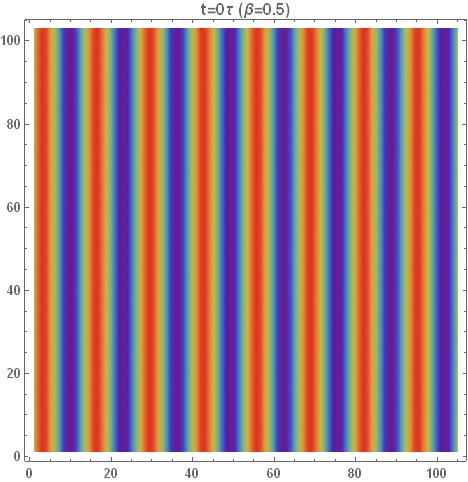}\includegraphics[scale=0.2325]{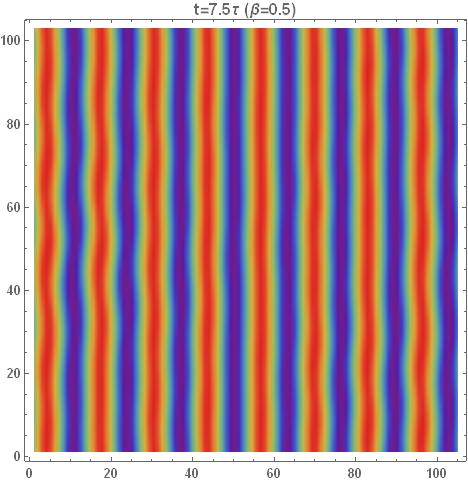}\includegraphics[scale=0.2325]{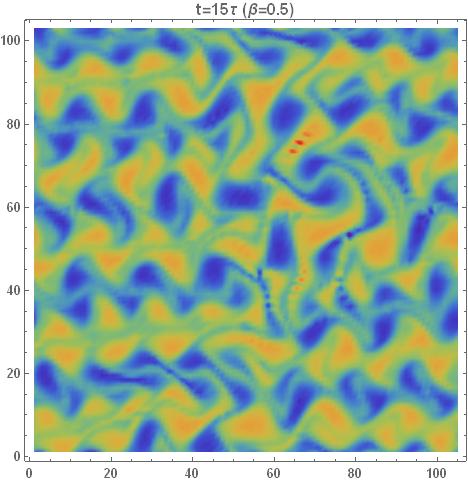}
        \includegraphics[scale=0.2325]{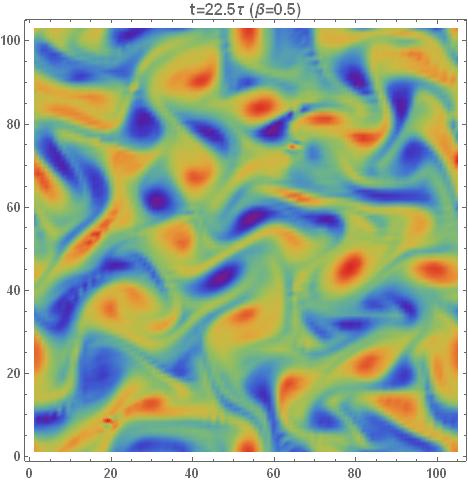}
        \includegraphics[scale=0.2325]{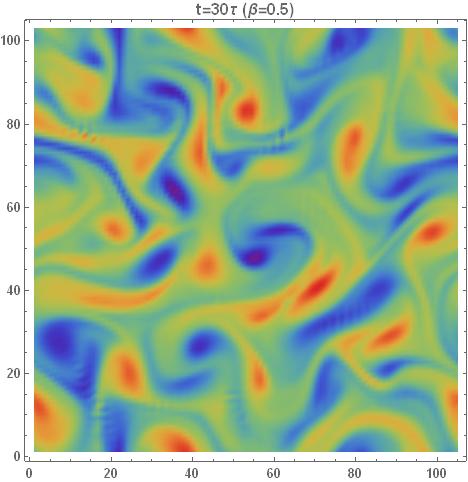}
        \includegraphics[scale=0.2325]{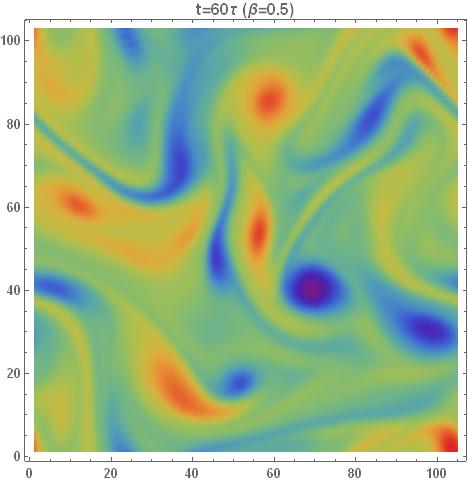}\includegraphics[scale=0.2325]{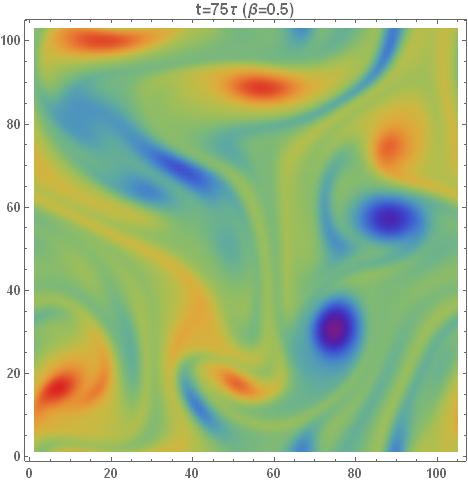}\includegraphics[scale=0.2325]{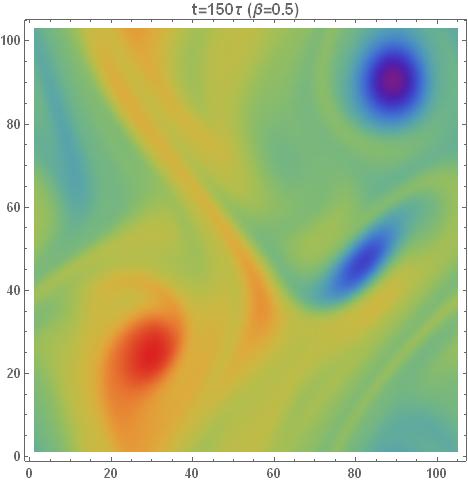}
        \caption{The vorticity field  of a two dimensional fluid flow in the EGB theory with $\beta=0.5$ and the other parameters are the same as those in Fig. \ref{fig:2Dvorticitybeta=1} at various times.}
        \label{fig:2Dvorticitybeta=05}
\end{figure}

\begin{figure}[hbtp]
        \centering
        \includegraphics[scale=0.2325]{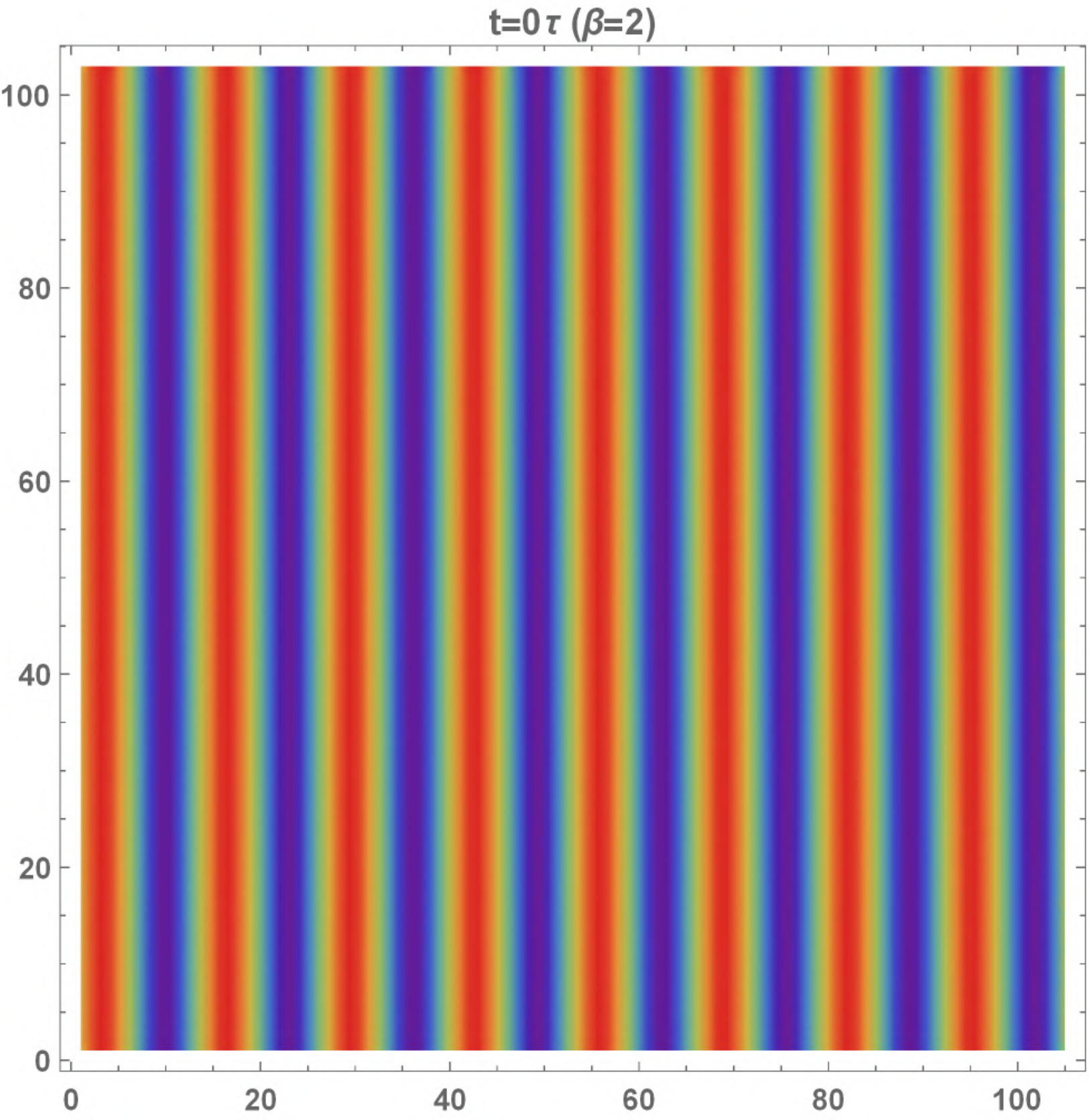}\includegraphics[scale=0.2325]{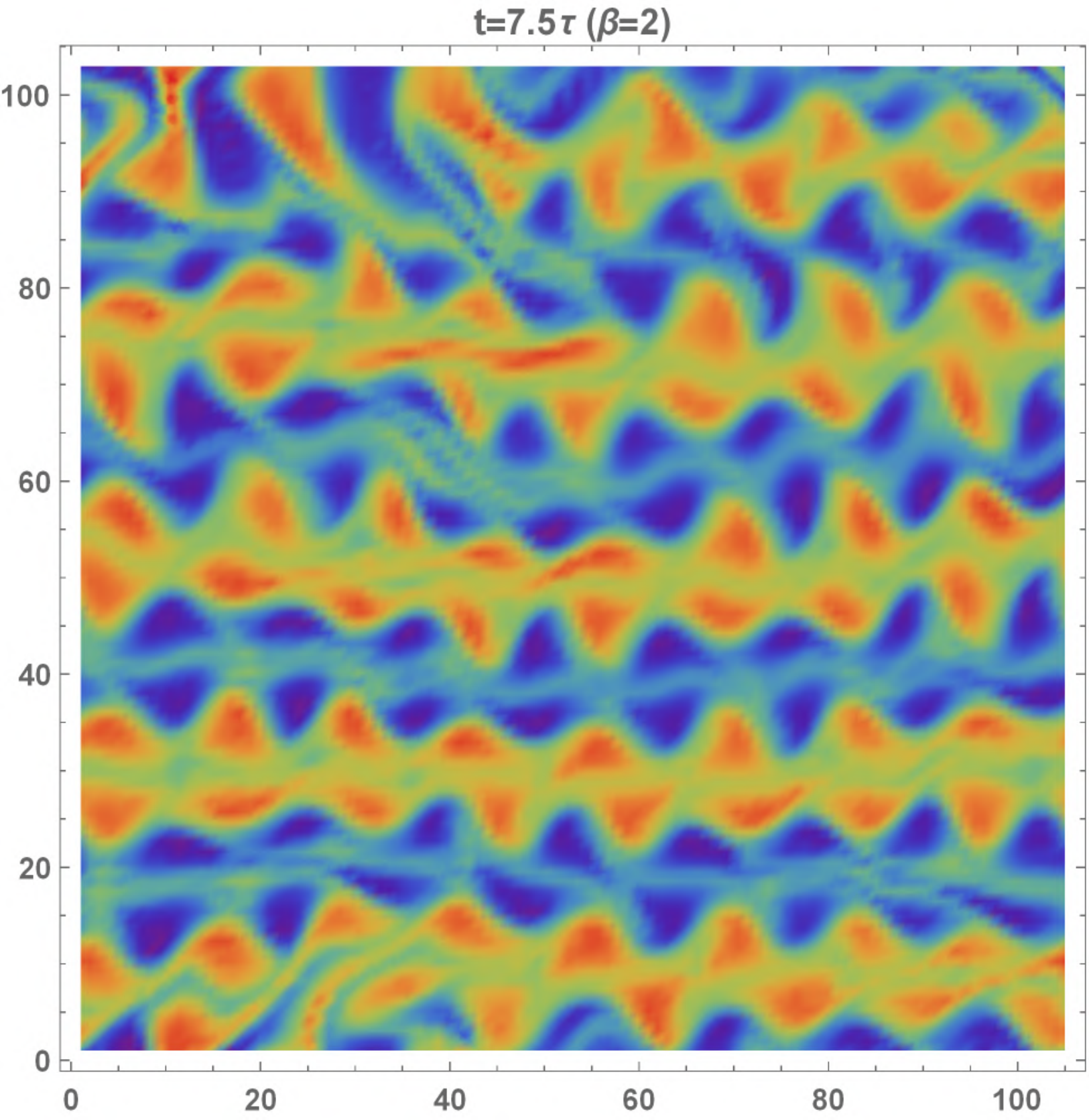}\includegraphics[scale=0.2325]{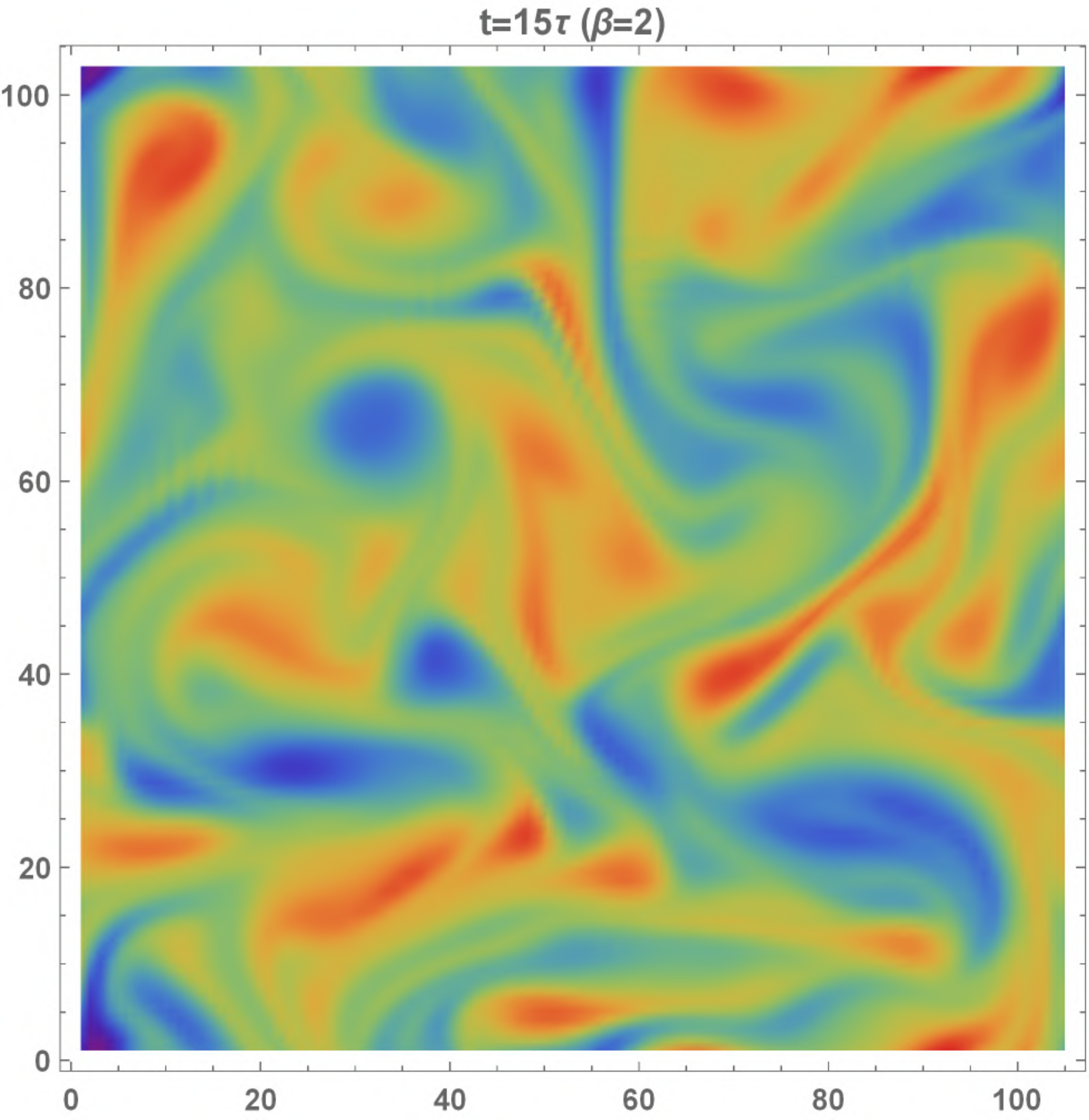}
        \includegraphics[scale=0.2325]{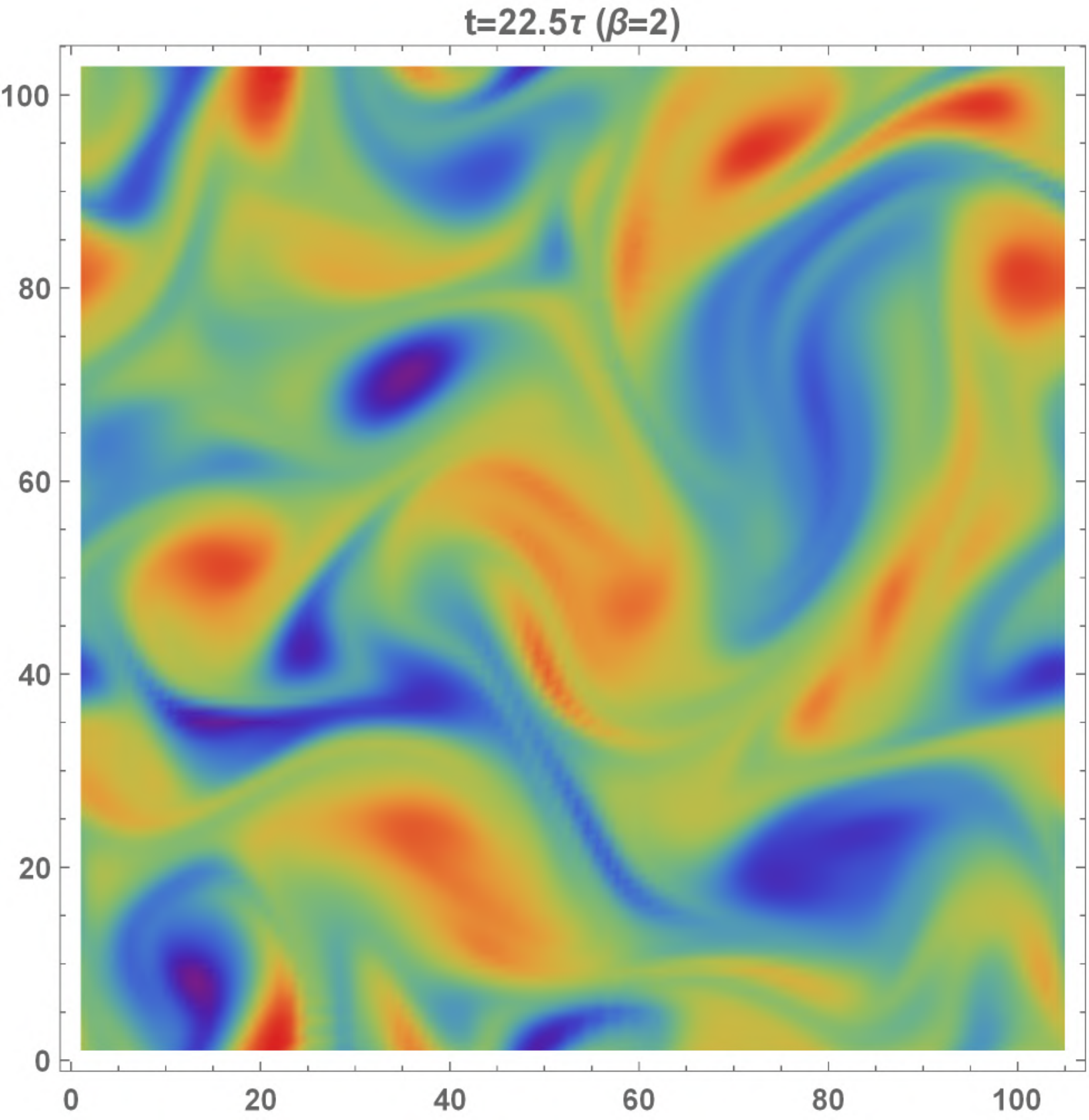}
        \includegraphics[scale=0.2325]{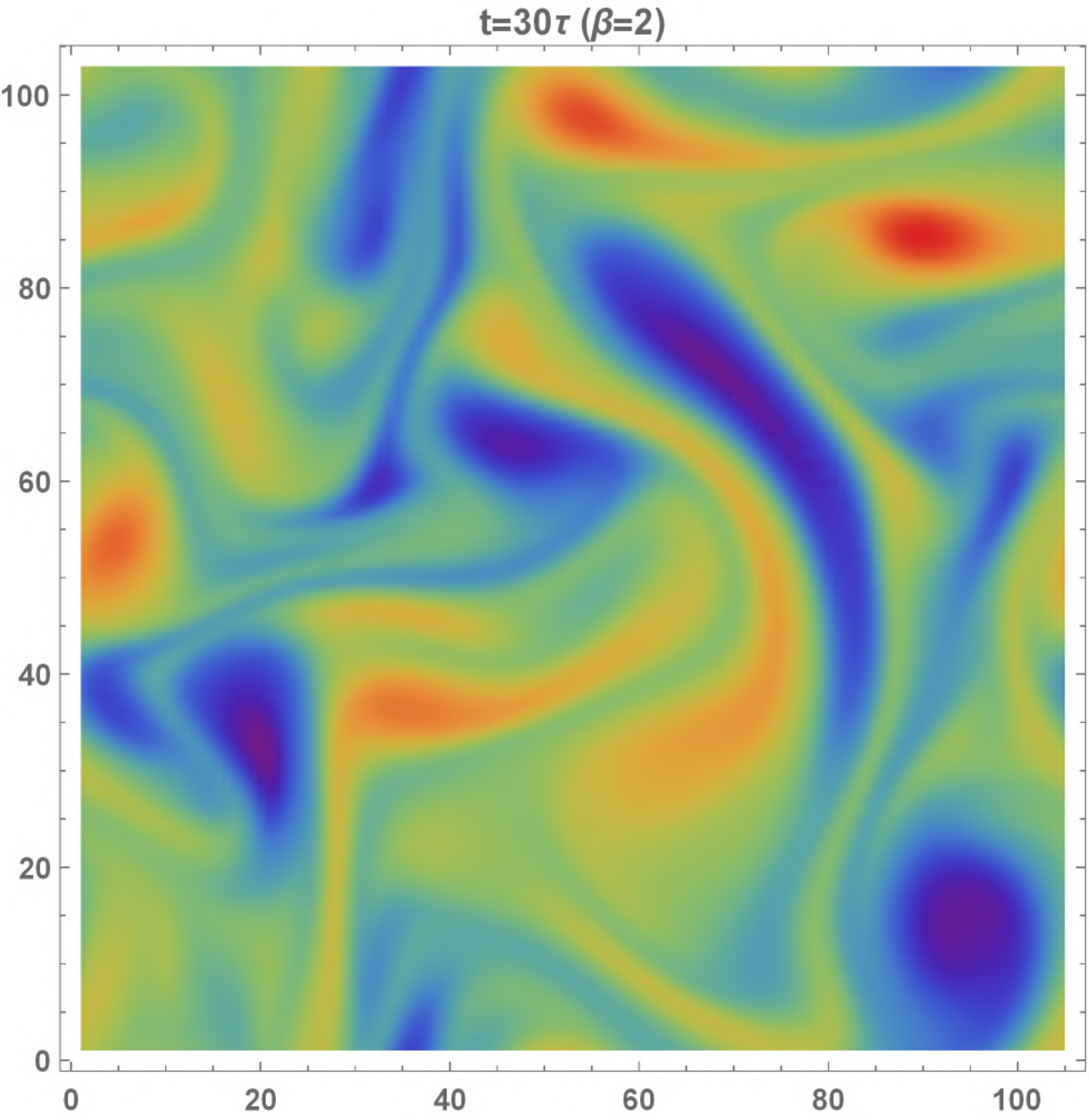}
        \includegraphics[scale=0.2325]{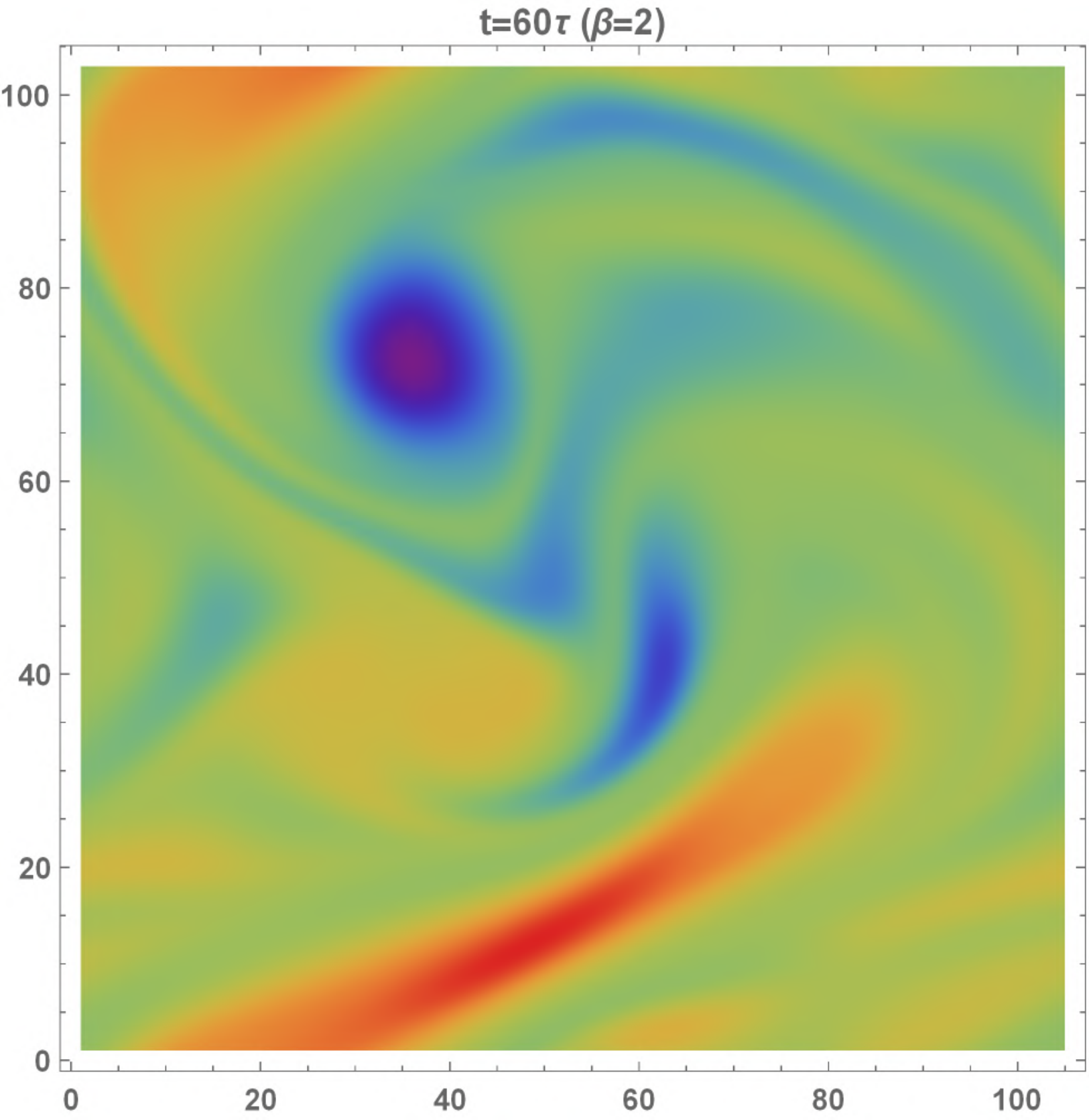}\includegraphics[scale=0.2325]{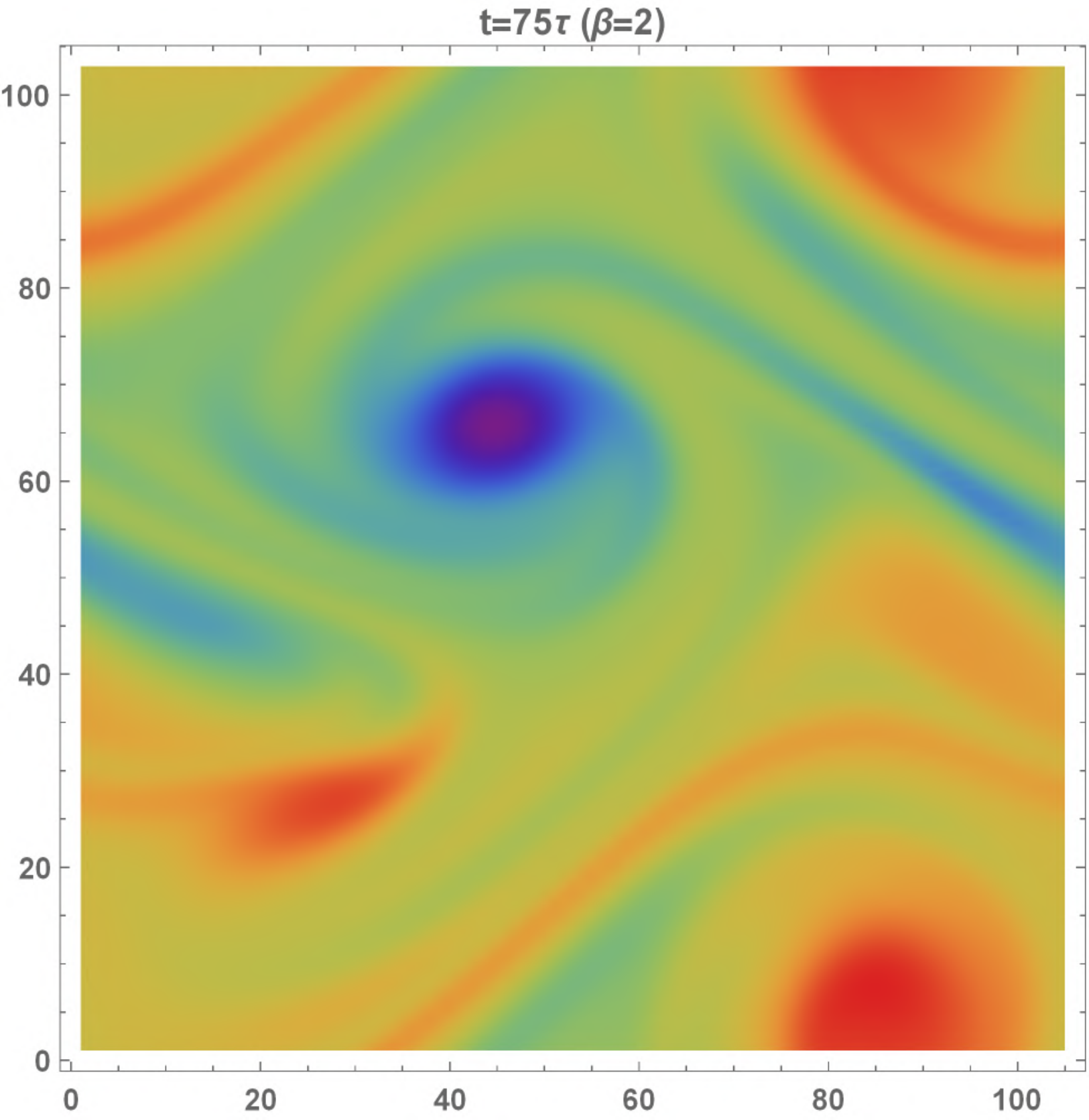}\includegraphics[scale=0.2325]{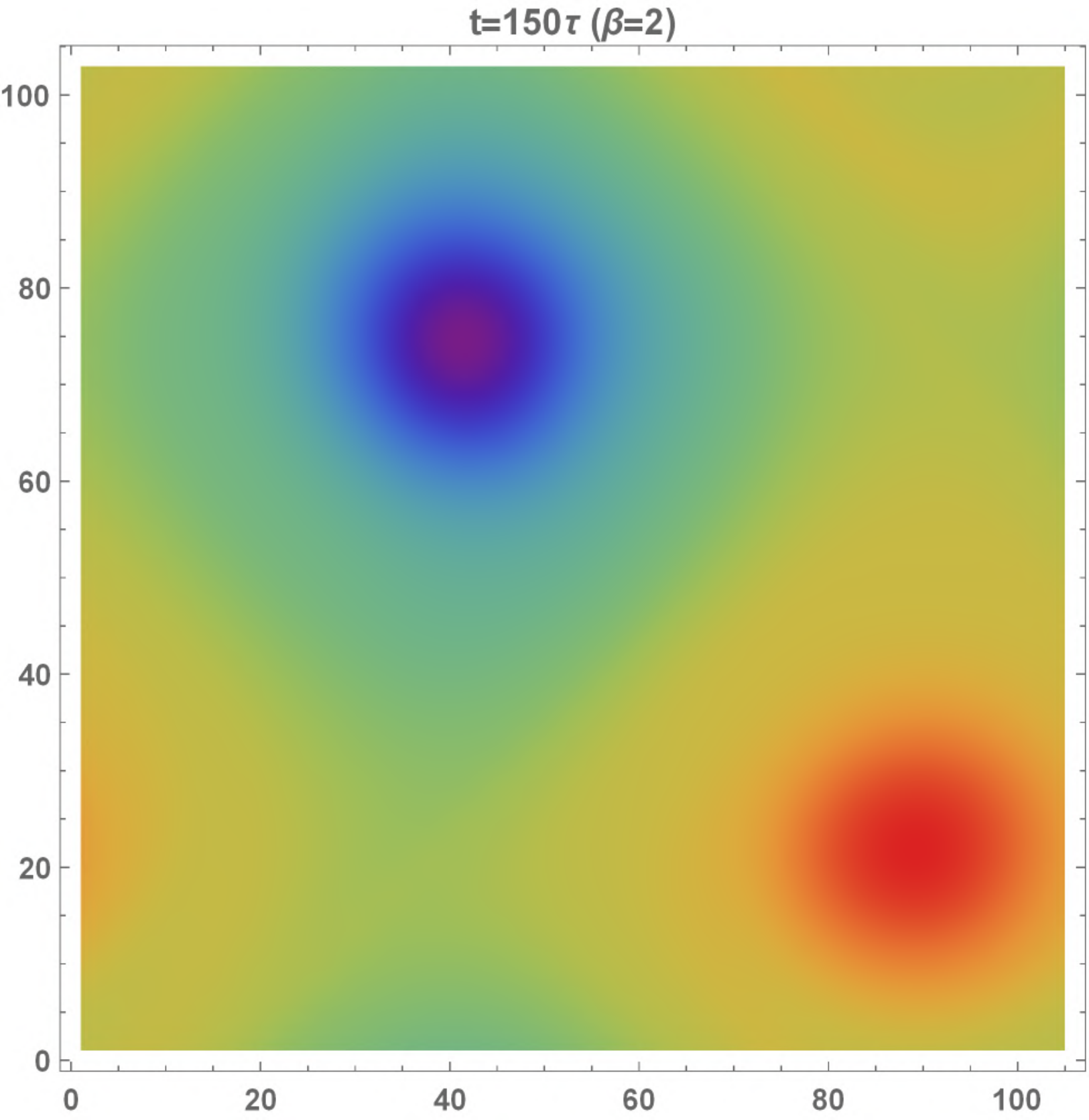}
        \caption{The vorticity field  of a two dimensional fluid flow in the EGB theory with $\beta=2$ and the other parameters are the same as those in Fig. \ref{fig:2Dvorticitybeta=1} at various times.}
        \label{fig:2Dvorticitybeta=2}
\end{figure}

\begin{figure}[hbtp]
        \centering
        \begin{tabular}{cccc}
      &\includegraphics[scale=0.255]{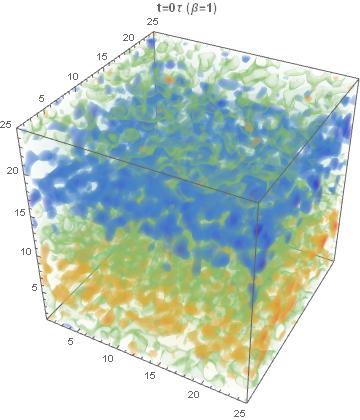}&\includegraphics[scale=0.255]{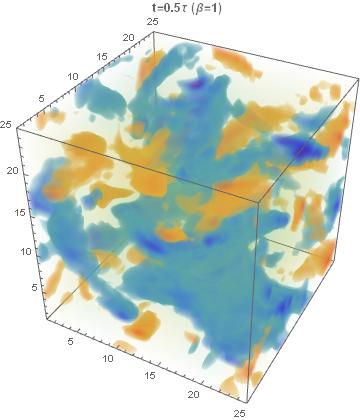}  &\includegraphics[scale=0.255]{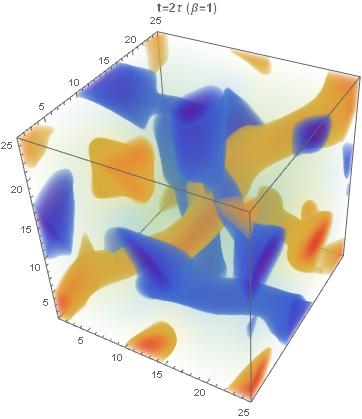}  \\
       & \includegraphics[scale=0.255]{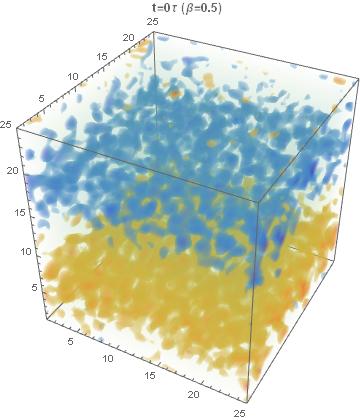}   & \includegraphics[scale=0.255]{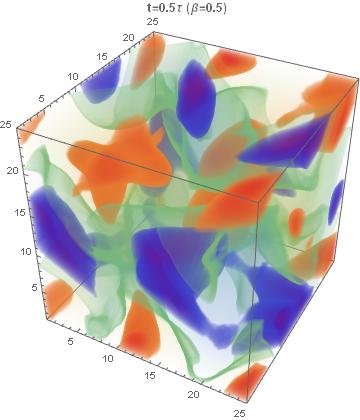}  & \includegraphics[scale=0.255]{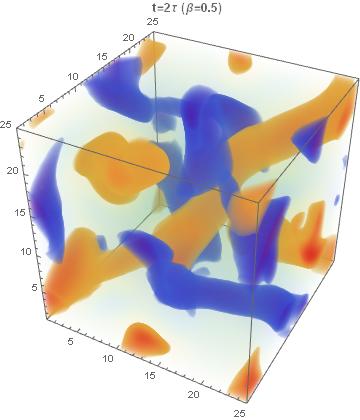}
        \end{tabular}
        \caption{The vorticity field $\vec{\omega} = \nabla\times\vec{p}$ of  three dimensional fluid flow with initial parameters $L=250$, $m_0=1$, $n_I=1$ and $A=10^{-2}$ at different times. For simplicity here we only show $\omega_z$, the other components are similar. The upper row is for the case of the Einstein gravity (with $\beta=1$) and the lower row is for the case of the EGB theory (with $\beta=0.5$). Different from the two dimensional flow, at the turbulent regime the formation of  small structure is observed which indicates the direct energy cascade.}
        \label{fig:vorticity3D}
\end{figure}

As a preview,  we show the typical simulations of two dimensional fluid flow in Figs. \ref{fig:2Dvorticitybeta=1}, Fig. \ref{fig:2Dvorticitybeta=05} and Fig. \ref{fig:2Dvorticitybeta=2} and the simulations of the three dimensional  fluid flow
in Fig. \ref{fig:vorticity3D} with large Reynolds numbers for different values of $\beta$. Qualitatively, the behavior of the fluid flows in the EGB theory is very similar to the one in the Einstein gravity. It  can be described by three phases: the initial growth of the instabilities, the turbulent regime in which the inverse/direct energy cascade is observed, and the late time decay into an equilibrium. An obvious distinction between these two theories is that, for a positive GB coefficient the time for the turbulence to emerge is later, but for a negative GB coefficient the time is earlier, under the same initial conditions ($L$, $m_0$ and $n_I$). This is largely because the presence of the positive GB coefficient lowers both the initial values of the Reynolds number and the Mach number as can be see from (\ref{initialRe}), thus it costs more time for the flow to reach the turbulent phase. In contrast, the presence of the negative GB coefficient lowers the initial value of the Reynolds number but increase the the initial value of the Mach number.
As a consequence, it costs less time for the flow to reach the turbulence phase.

\begin{figure}[hbtp]
        \centering
        \includegraphics[scale=0.3]{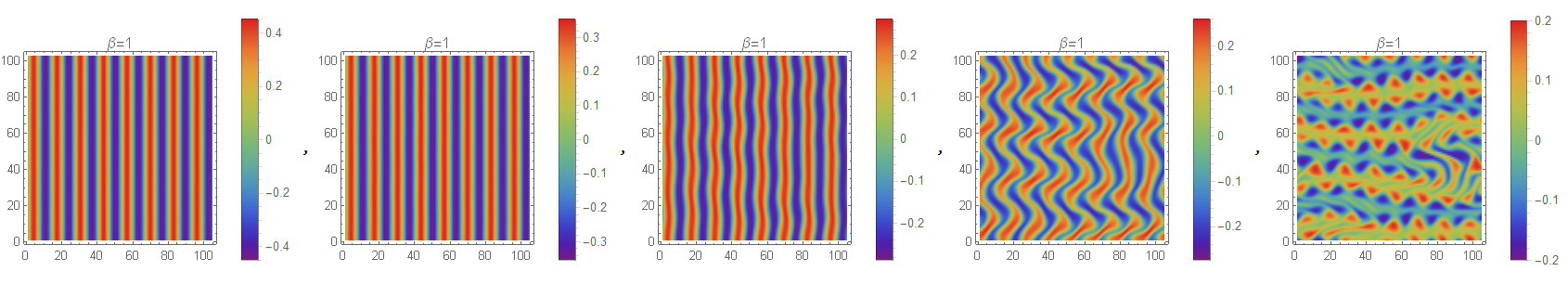}
        \includegraphics[scale=0.3]{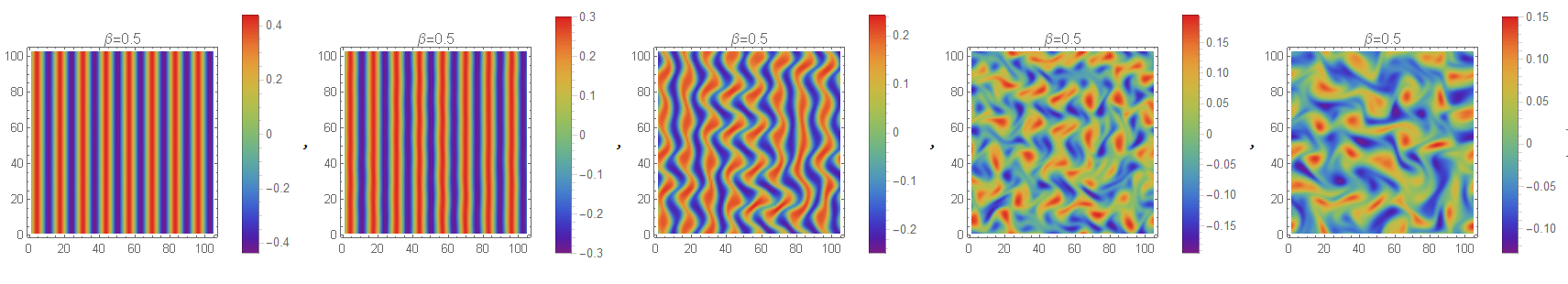}
        \includegraphics[scale=0.3]{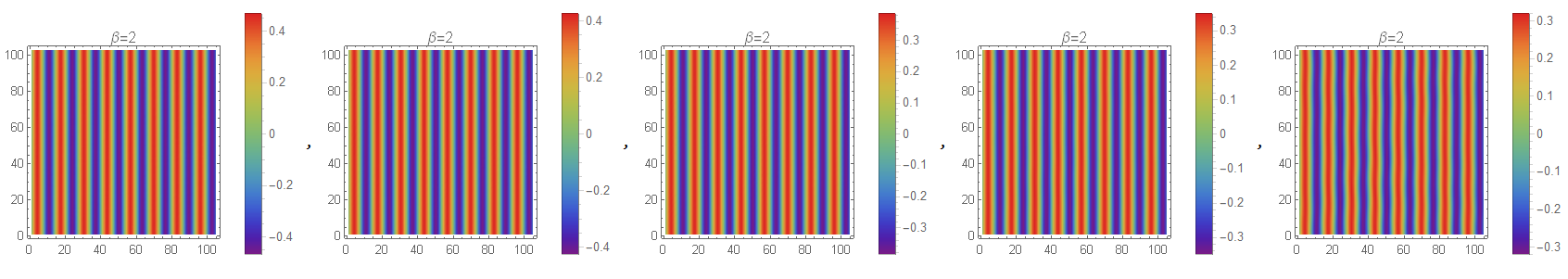}
        \caption{The vorticity field of two dimensional fluid flow in the EGB theory with various values of $\beta$. Here by adjusting $L$ and $m_0$ we keep the initial Reynolds number and Mach number invariant with $\beta$.}
        \label{fig:2DRe500}
\end{figure}

 On the other hand, if we keep the initial Reynolds number and Mach number fixed, then the behavior of the fluid for various $\beta$'s should not be the same, because their dynamical equations are different. As shown in Fig. \ref{fig:2DRe500}, we can see that in comparison with the case of the Einstein gravity, a positive GB coefficient has a larger evolution rate while a  negative GB coefficient has a smaller evolution rate. This may be ascribed to the distinction of the dynamical equation. After we adopt the definition (\ref{ReandM}), the dynamical equation
has  two extra terms proportional to the inverse of the Reynolds number. These two terms induce viscous forces and affect the evolution of the fluid flow. As we will see in the next subsection, from the linear analysis, the positive GB coefficient has the smaller viscous effect and the larger evolution rate, while the negative GB coefficient has the larger viscous effect and the smaller evolution rate.

 In the following  we will pay our attention to the effects of the GB term on the fluid flows in these different stages.


\subsection{Stability of shear flow}
In this subsection we analyze the early stage of the flow. Due to the restriction of our computational resource, in the following we mainly show our results of the two dimensional flows. As we mentioned before, depending on the initial Reynolds number, the  small random perturbations of the shear flow grow or decay with time. The growth of the perturbations can be quantified by the
growth of $p_x$ field \cite{Green1309} or as  in \cite{Rozali1707} by the energy spectrum $E_C$ whose definition is
\footnote{The definition of the energy spectrum is a little different from the one in \cite{Rozali1707}: here we use $\vec{p}$ rather than $\vec{u}$ in $E_C$. We find
that this would be easier for numerical analysis. It works pretty well. }
\be\label{energyspectrum}
E_C(v,k) = \frac{\partial}{\partial k}\int_{|\vec{k}'|\leq k}\frac{d^{q}k'}{(2\pi)^{q}}\Big|\vec{p}_q(v,\vec{k}')\Big|^{2},
\ee
with
\be
\vec{p}_q(v,\vec{k}) = \int d^{q}x\,\vec{p}(v,\vec{x})e^{-i\vec{k}\cdot\vec{x}}.
\ee
 In what follows we will take either one of the two quantities when necessary.

\begin{figure}[hbtp]
        \centering
        \includegraphics[scale=0.225]{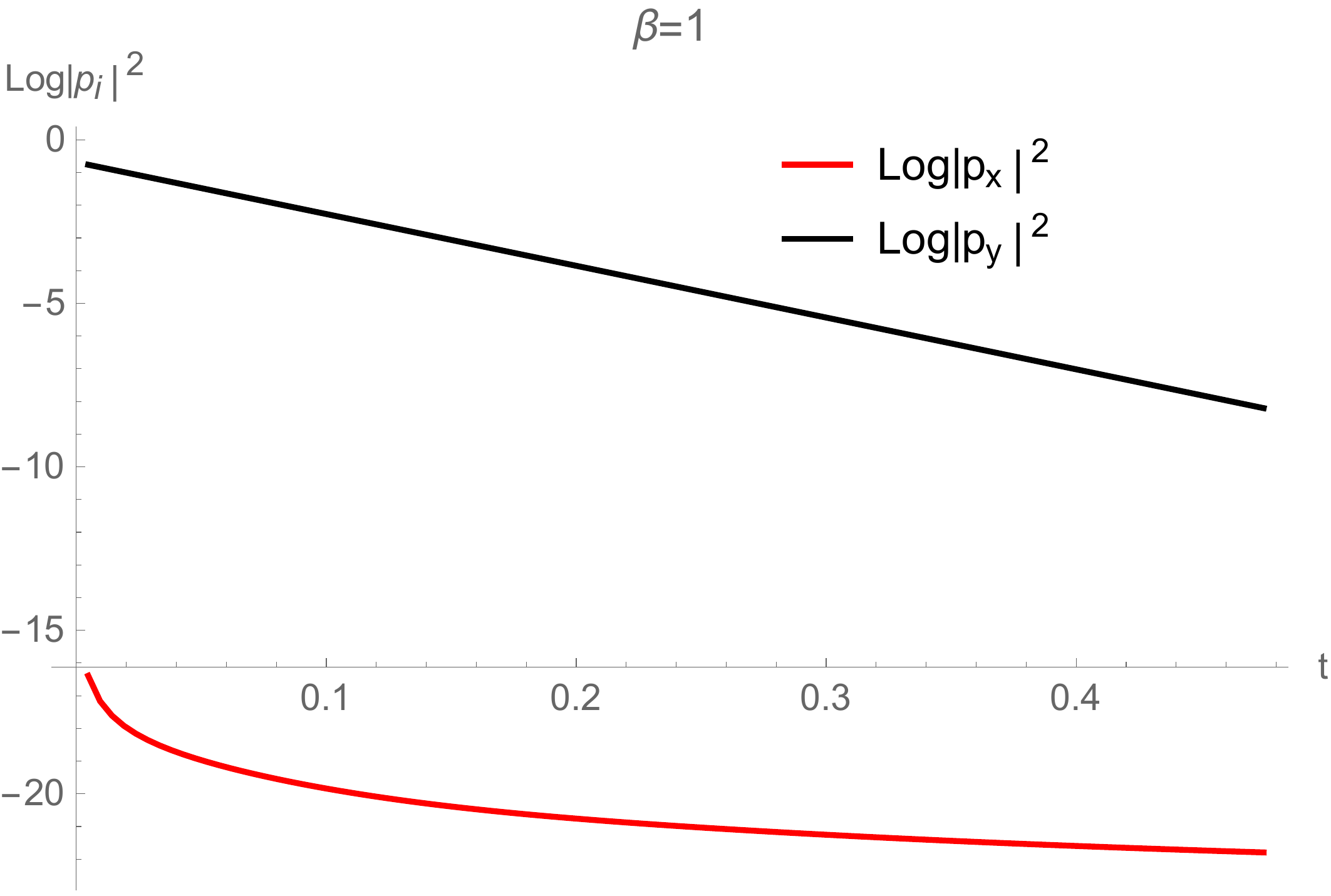}
        \includegraphics[scale=0.225]{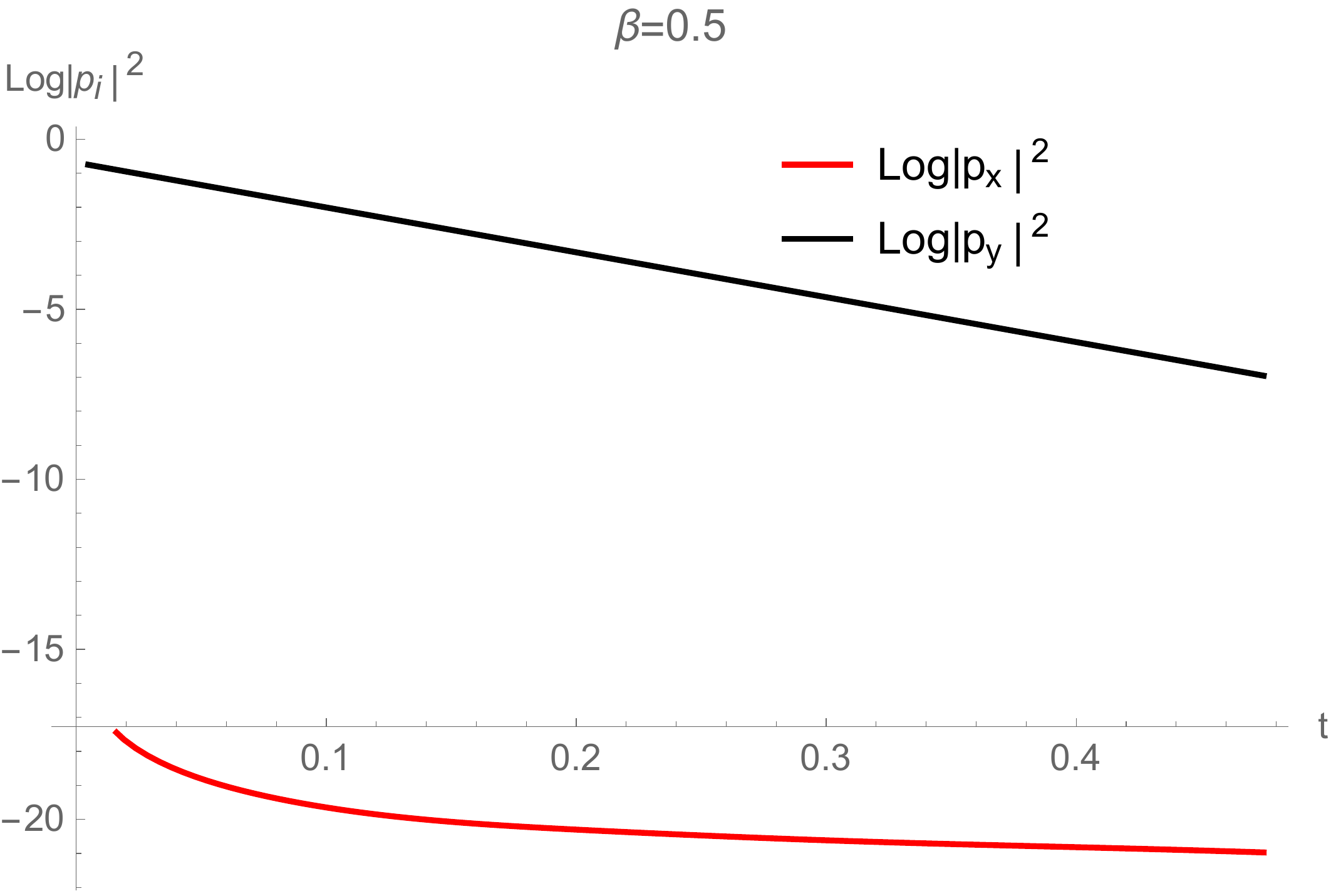}
        \includegraphics[scale=0.225]{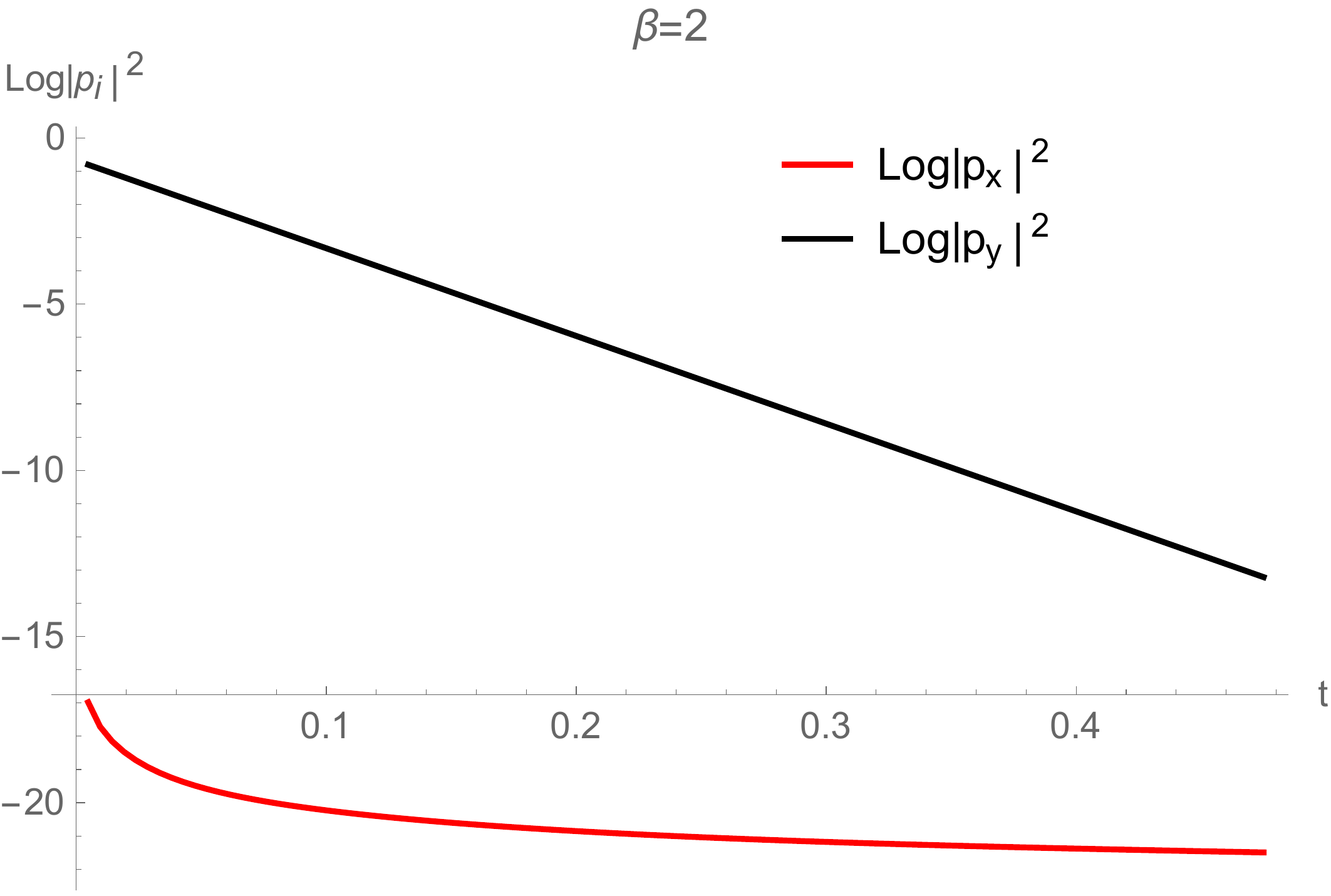}
        \caption{$L_2$ norms of $p_i$ as a function of time for a laminar flow, with $L=50$, $n_I=5$ and $m_0=2$.  }
        \label{fig:2DinitialstageRe10}
\end{figure}

At low values of  $\bRe_0$ the flow is laminar, as shown in Fig. \ref{fig:2DinitialstageRe10}. We can see that the $L_2$ norms of the components of $p_i$,
\be
L_2(p_i)=\frac{\int p_i^2d^{q}x}{\int d^{q}x},
\ee
  decay exponentially both for the  Einstein gravity and the EGB theory. From the linear analysis in section \ref{subsection:linearanalysis}, we know that the presence of  the positive GB coefficient lowers the quasinormal mode frequency (\ref{QNMfrequency}) and so the viscosity (\ref{viscosity}), thus the  $L_2$ norms of $p_i$ decay slower than the ones in the Einstein gravity. In contrast, the presence of the negative  GB coefficient increases the decay rate so that we have steeper lines in the right panel of Fig. \ref{fig:2DinitialstageRe10}. Moreover, we find that for larger Mach numbers the results are essentially the same as the ones of small Mach number case.

\begin{figure}[hbtp]
        \centering
        \includegraphics[scale=0.3]{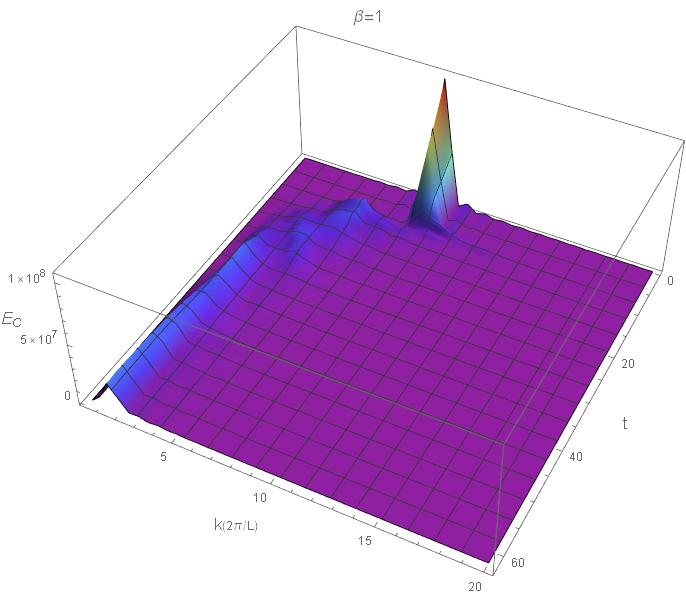}
        \includegraphics[scale=0.3]{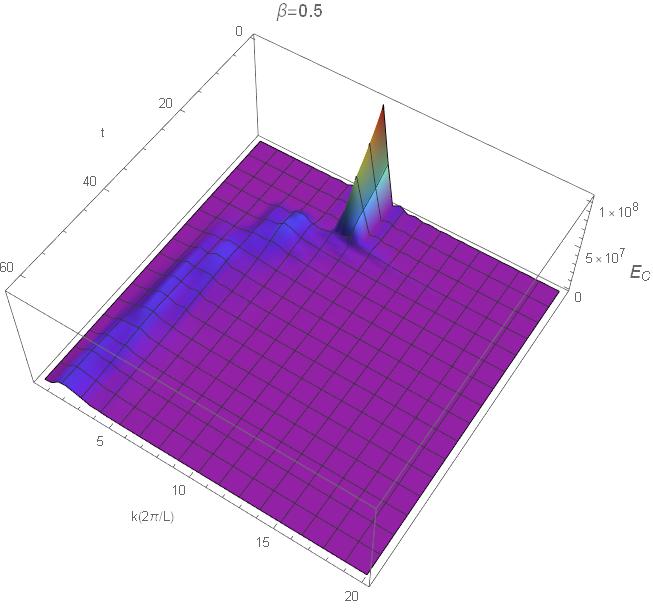}
        \caption{The energy spectrum $E_C$ for two dimensional fluid flow, with $L=15000$, $n_I=8$ and $m_0=2$. The left panel corresponds to $\beta=1$ and the right
        panel corresponds to $\beta=0.5$.}
        \label{fig:2DinitialstageRe937}
\end{figure}
 \begin{figure}[hbtp]
        \centering
        \includegraphics[scale=0.225]{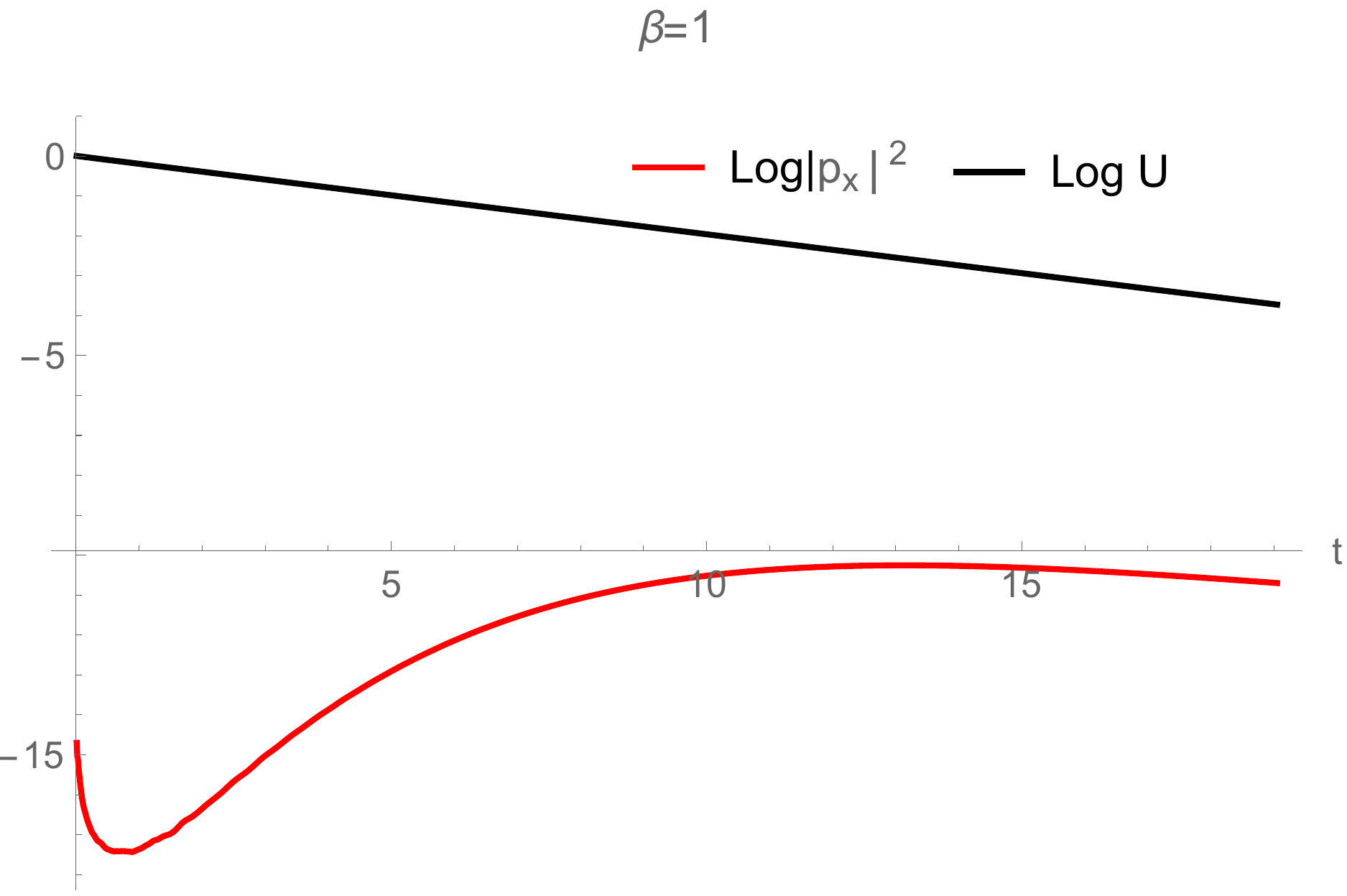}
        \includegraphics[scale=0.225]{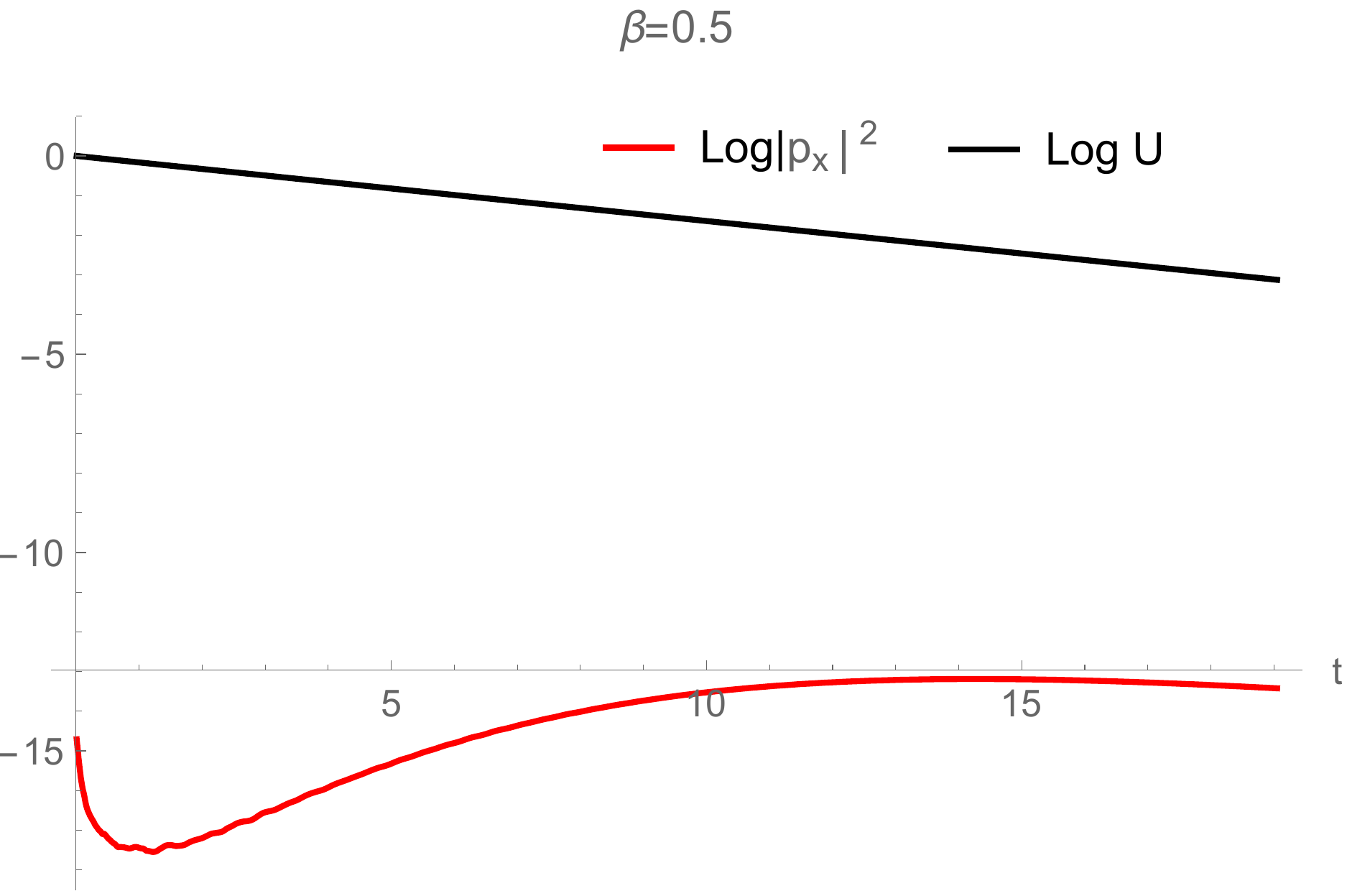}
        \includegraphics[scale=0.225]{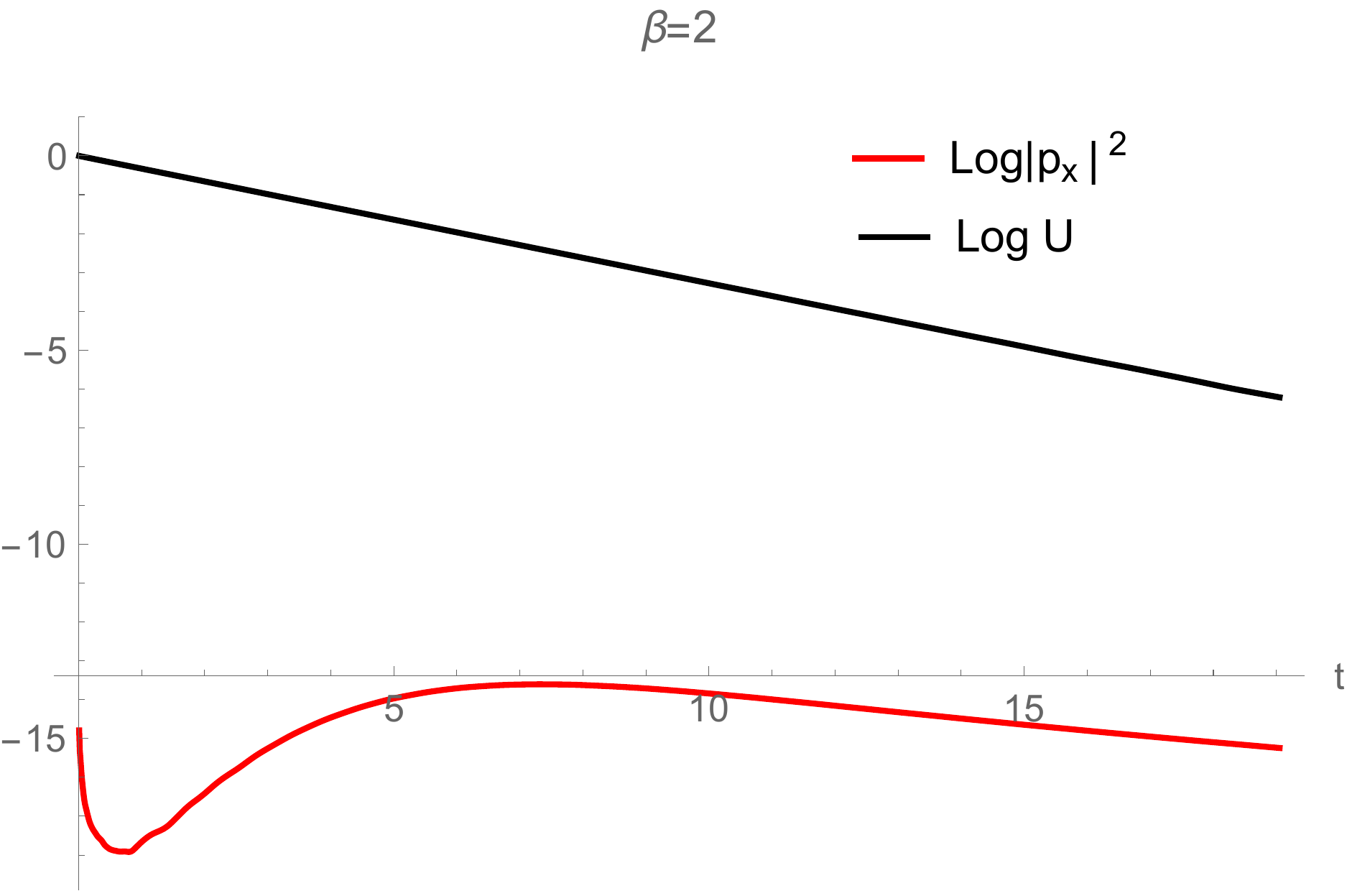}
        \caption{Intermediate flow:  $L_2(p_x)$ and the characteristic velocity $U$ as  the functions of time for a laminar flow, with  $L=2000$, $n_I=5$ and $m_0=2$.
        For the positive GB coefficient ($\beta=0.5$), the decay rate of $\log U$ is smaller and the maximum of  $\log L_2(p_x)$ appears at later time. In contrast, for the
        negative  GB coefficient ($\beta=2$), the decay rate is larger and it takes less time for $\log L_2(p_x)$ to reach the peak. }
        \label{fig:2DcriticalRe}
\end{figure}

At a large enough $\bRe_0$ we observe the turbulent instabilities, as shown in Fig. \ref{fig:2DinitialstageRe937}. At the first glance the energy spectrums in the Einstein gravity and the EGB theory are very similar. In both cases we find the decay of the initial disturbance at $k=8\times\frac{2\pi}{L}$ and then an unstable mode emerging at a lower wavenumber $k\sim \frac{3}{2}\times\frac{2\pi}{L}$, which may indicate the appearance of the inverse cascade. However, the direction cascade is not apparent, this may attribute the insufficient precision of our numerical simulation. As we mentioned before since the flow in the EGB theory has a lower Reynolds number,  the initial disturbance decays slower and  the new unstable mode grows slower, comparing  with the  Einstein gravity.

For an intermediate value of the initial Reynolds number, as shown in Fig. \ref{fig:2DcriticalRe},   the initial small perturbations characterized by
$L_2(p_x)$ grow exponentially to reach a maximum that is smaller than $L_2(p_y)$, and then decay exponentially. From Fig. \ref{fig:2DcriticalRe} we can
 find the exponential decay of the characteristic velocity $U$ defined in (\ref{characteristicvelocity}). From the definition (\ref{ReandM})  the Reynolds number
 varies with time. In this case  we can define a critical Reynolds number $R_c$ as in \cite{Green1309}. Initially, we have $\bRe_0>R_c$, thus the unstable mode $L_2(p_x)$ grows exponentially. As time progresses, $\bRe $ decreases,  then when $\bRe < R_c$, $L_2(p_x)$ is not growing any more, instead it decays exponentially. Therefore, at the peak of $L_2(p_x)$ we may identify $\bRe=R_c$.

 \begin{table}[h]
 \centering
\begin{tabular}{|c|c|c|c|c|}
\hline
$\bRe_0$&$150$&$160$&$170$&$180$\\\hline
$R_c$&$17.45$&$17.46$&$18.82$&$20.68$\\\hline
\end{tabular}
\caption{The critical Reynolds numbers of the fluid flow in the Einstein gravity with the initial Mach number $\bM_0=0.5$.}\label{Rc:fixedMach}
\end{table}%
\begin{table}[h]
 \centering
\begin{tabular}{|c|c|c|c|c|}
\hline
$\bM_0$&$0.25$&$0.5$&$1$&$2$\\\hline
$R_c$&$18.11$&$17.46$&$16.30$&$15.16$\\\hline
\end{tabular}
\caption{The critical Reynolds numbers of the fluid flow in the Einstein gravity with the initial Reynolds number $\bRe_0=160$.}\label{Rc:fixedReynolds}
\end{table}
\begin{table}[h]
 \centering
 $\bM_0=1$\quad
\begin{tabular}{|c|c|c|c|c|}
\hline
$\beta$&$1/3$&$0.5$&$1$&$2$\\\hline
$R_c$&$16.42$&$16.24$&$16.30$&$16.29$\\\hline
\end{tabular}
 \\$\bM_0=2$\quad
\begin{tabular}{|c|c|c|c|c|}
\hline
$\beta$&$1/3$&$0.5$&$1$&$2$\\\hline
$R_c$&$15.00$&$15.05$&$15.16$&$15.13$\\\hline
\end{tabular}
\caption{The critical Reynolds numbers of the fluid flow in the EGB gravity with different GB coefficients, here the initial Reynolds number is fixed $\bRe_0=160$.}
\label{Rc:differentGBcoe}
\end{table}

We have identified the critical Reynolds number $\bRe_c$ for different initial Reynolds numbers, different initial Mach numbers and different $\beta$'s. Firstly, for the Einstein gravity $\beta=1$, we find that for a given Mach number the
 larger is initial Reynolds number, the  larger is the critical Reynolds number, as shown in Table \ref{Rc:fixedMach}. Secondly, for a given initial Reynolds number, we find that the larger
is the Mach number,  the smaller is the critical Reynolds number, as shown in Table \ref{Rc:fixedReynolds}.
 Finally, we can see from Table \ref{Rc:differentGBcoe} that for the fixed initial Reynolds number and Mach number, the critical Reynolds numbers of
 different GB coefficients are pretty close.
\subsection{Turbulent regime}
\begin{figure}[hbtp]
        \centering
        \includegraphics[scale=0.325]{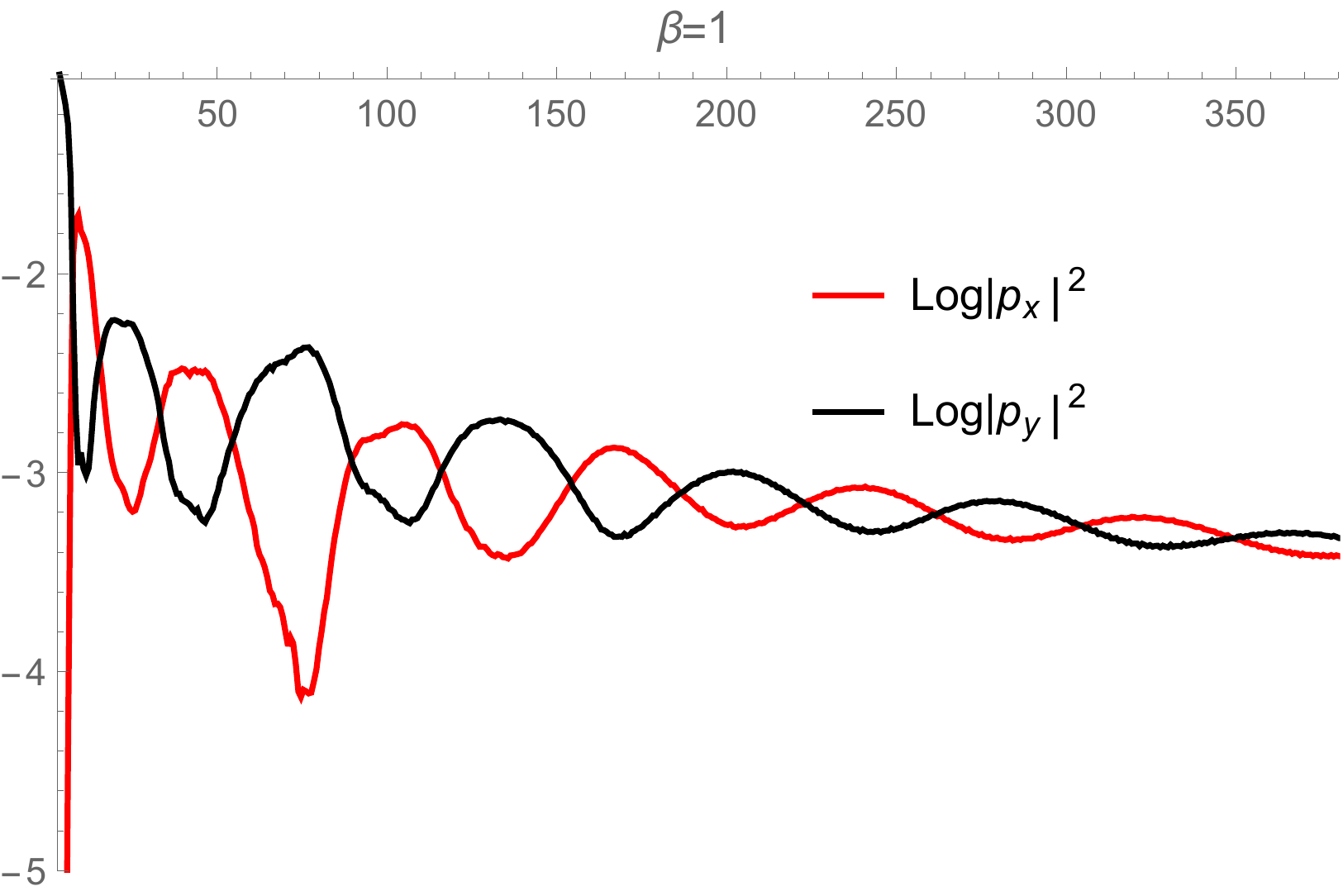}
        $\qquad$
        \includegraphics[scale=0.325]{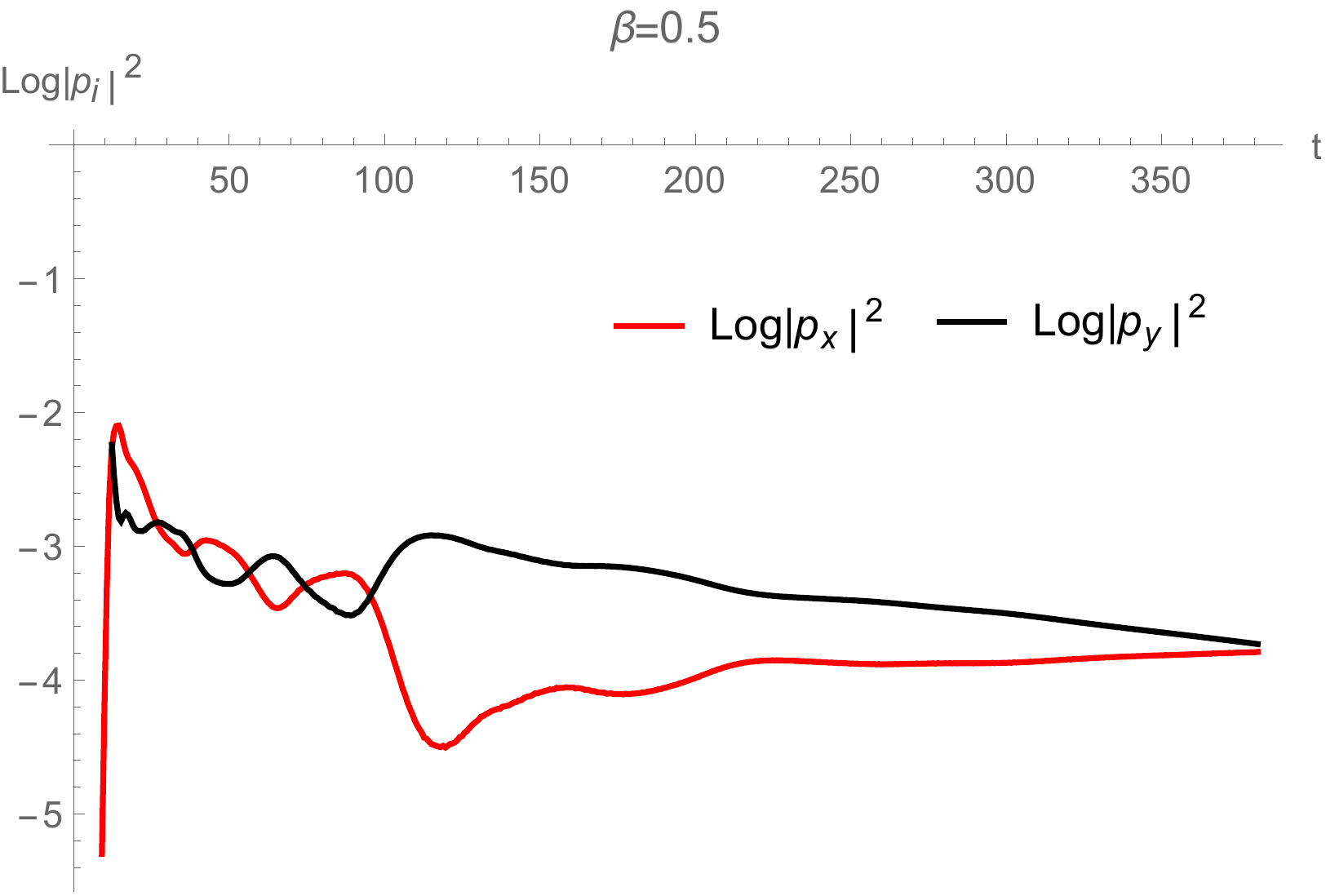}
        \caption{The $L_2$ norm of $p_i$ as a function of time for the turbulent flow, with $L=15000$, $n_I=8$ and $m_0=2$. We can see that $p_x$ grows exponentially fast until it is of similar amplitude to $p_y$.}
        \label{fig:2DpxpyRe937}
\end{figure}
\begin{figure}[hbtp]
        \centering
        \includegraphics[scale=0.445]{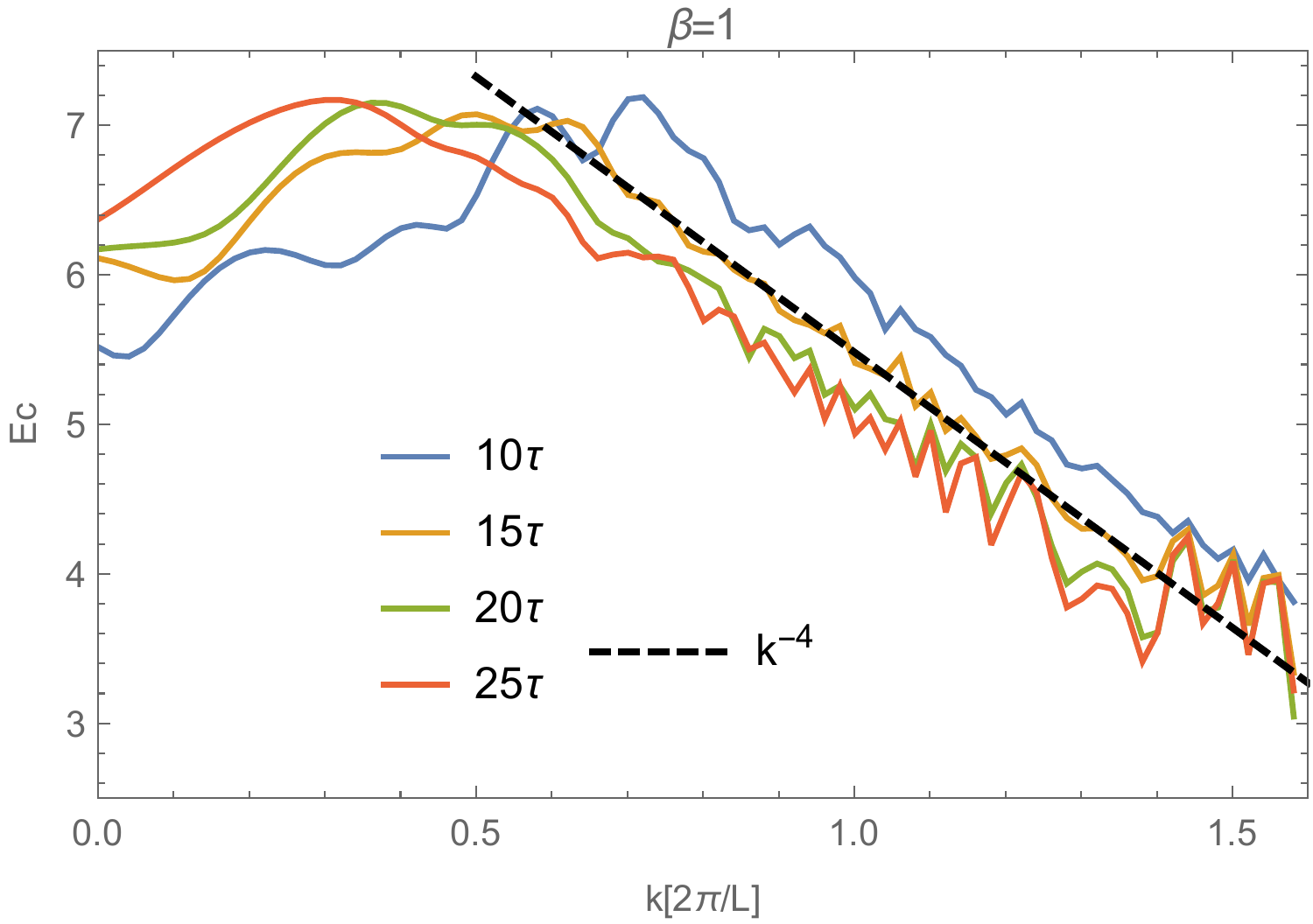}
        \includegraphics[scale=0.425]{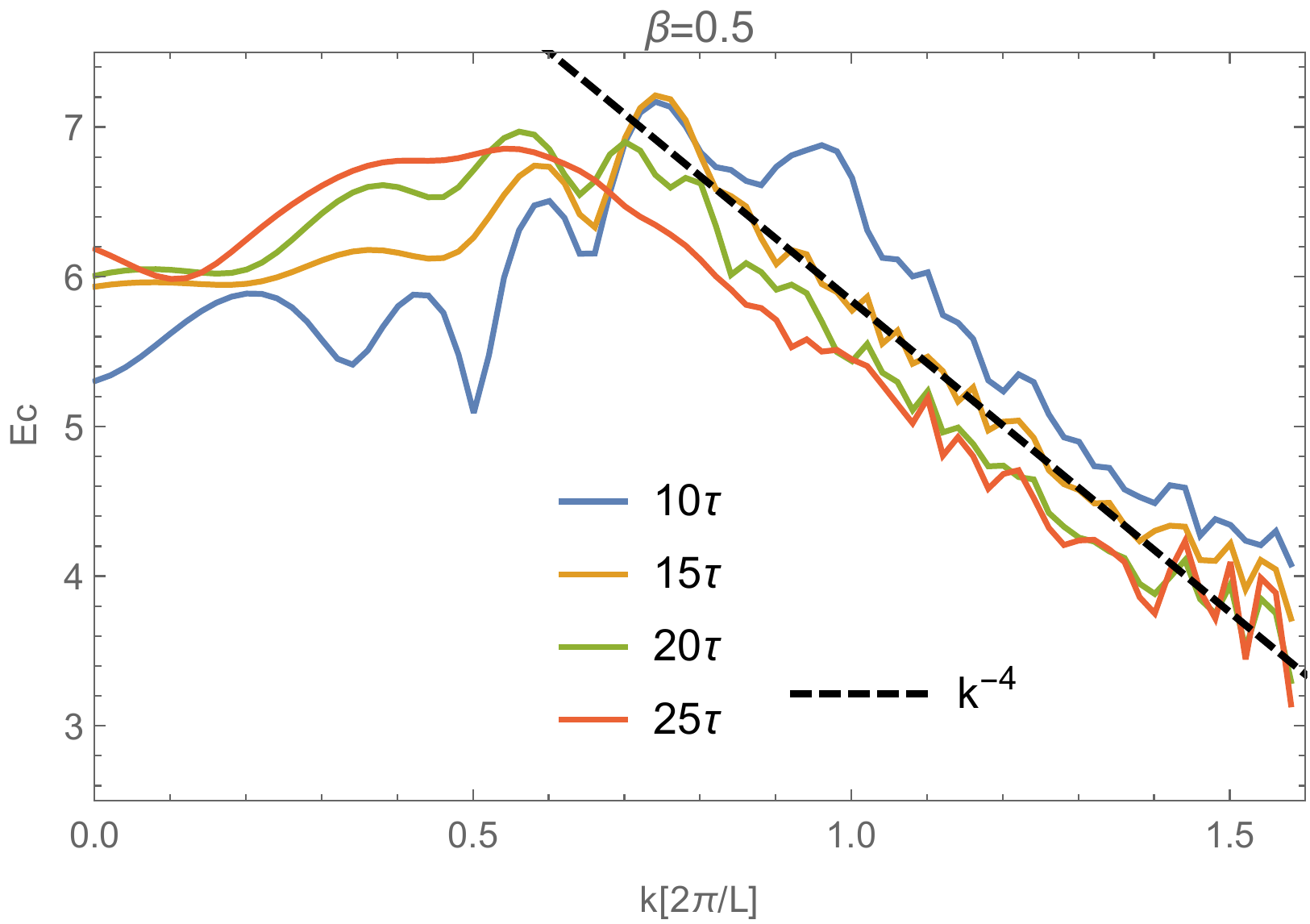}
        \caption{The  log-log plots of the energy spectrum for two dimensional flow at various times, with $L=15000$, $n_I=8$ and $m_0=2$.
        There appears an power law $E_C\sim k^{-4}$ in the inertial range. }
        \label{fig:2Dpowerlawbeta}
\end{figure}
If the initial Reynolds number is sufficiently large, the initial instability grows until it reach the same amplitude as that of the initial shear mode, then the fluid goes into the turbulent regime, as shown in Fig. \ref{fig:2DpxpyRe937}. In this stage, the small eddies merge into vortices, which continue to merge into increasingly large vortices until two counter-rotating vortices are formed, as shown in Figs. \ref{fig:2Dvorticitybeta=1} and \ref{fig:2Dvorticitybeta=05}. This process is the so-call inverse energy cascade.
 Besides, there is a direct cascade caused by the formation of the filaments of the vorticity. These two cascades can be characterized by the shape of the energy spectrum $E_C$ in the inertial range. The detailed analysis of the  energy spectrum during the turbulent phase can be found in \cite{ Rozali1707}. Here we only want to emphasize that for the two dimensional fluid  flow in
the EGB theory the  energy spectrum of the direct cascade satisfies a $k^{-4}$ power law for a small Mach number and a large Reynolds number, as shown in Fig. \ref{fig:2Dpowerlawbeta}. Similar to the Einstein gravity, the $k^{-5/3}$ law for the inverse cascade is absent from
the energy spectrum analysis in the case of the EGB theory. From the analysis in \cite{Mininni2013}, it was found that the range at which the $k^{-5/3}$ law applies is very short
  and the distinction from the $k^{-3}$ law demands high numerical precision.

For three dimensional fluid flow,   the celebrated $k^{-5/3}$ power law has been found in \cite{Rozali1707}, even the Mach number is large. Due to the limited computational power, we are not able to carry out the numerical analysis on three dimensional fluid flow. Nevertheless,
we expect the same energy cascade happens in the EGB turbulence as well. 

\subsection{Late time decay}
\begin{figure}[hbtp]
        \centering
        \includegraphics[scale=0.325]{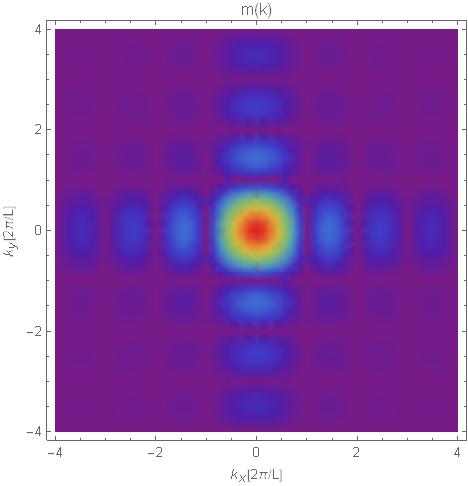}\includegraphics[scale=0.325]{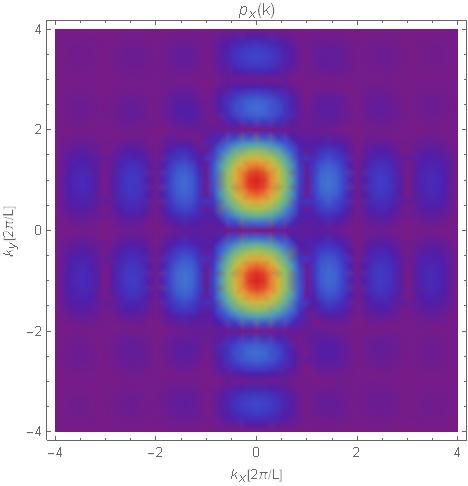}\includegraphics[scale=0.325]{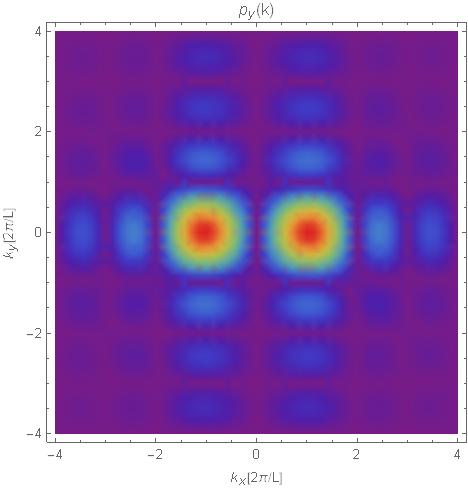}
        \caption{The energy density and the velocity field  at the final stage of the two dimensional turbulent flow. }
        \label{fig:2Dlatetimedecaykspace}
\end{figure}
The final phase of the turbulent flow is the formation of two counter-rotating vortices,  as shown in Figs. \ref{fig:2Dvorticitybeta=1}, \ref{fig:2Dvorticitybeta=05} and  \ref{fig:2Dvorticitybeta=2}.
We find that the late time decay takes the following form
\bea
m(x,y)&=&m_0,\nonumber\\
p_x(x,y)&=& p_{x0}\, e^{-\frac{\beta^2+1}{\beta+1} \big(\frac{2\pi}{L}\big)^2v} \cos\Big(\frac{2\pi}{L}y+\phi_1\Big),\nonumber\\
p_y(x,y)&=& p_{y0}\, e^{-\frac{\beta^2+1}{\beta+1} \big(\frac{2\pi}{L}\big)^2 v} \cos\Big(\frac{2\pi}{L}x+\phi_1\Big),
\eea
where $m_0$, $p_{x0}$ and $p_{y0}$ are constant. The above behavior can be understood as follows. Since in this stage the velocity field becomes small, thus the linear analysis for the fluid flows is applicable. The decay rate is just the damping part of the quasinormal mode of the black brane. On the other hand, due to the inverse cascade at late time the lowest mode dominates the flow, as shown in Fig. \ref{fig:2Dlatetimedecaykspace}.

\section{Geometric Interpretation of Turbulence}\label{section:geometricinter}
In this section, we would like  to see if the horizon power spectrum defined in \cite{Adams1307} is  applicable to the
turbulence of the holographic EGB fluid.

First of all, let us write down the leading order metric of the EGB-AdS black brane
\be\label{leadingordermetric}
ds^2=-\frac{1-\mathbf{b}}{2\tilde{\alpha}} dv^2+2dvdr-\frac{1}{n}\frac{ \beta (\bb-\beta) p_a}{\tilde{\alpha}\, m} dz^a dv+\frac{r^2}{n}\Big(\delta_{ab}+\frac{G_0}{n}\Big)dz^adz^b +\frac{r^2}{n}d\vec{x}^2,
\ee
where $\bb$ and $\beta$ are introduced in (\ref{bandbeta}), and $G_0$ is given in (\ref{G0}).
Then the event horizon is at
\be\label{horizonposition}
\sR=m(v,z^a).
\ee
The extrinsic curvature on the event horizon is given by
\be
\Theta_{\mu\nu}\equiv \Pi^\rho_{\;\mu} \Pi^\sigma_{\;\nu}\nabla_{\rho} n_\sigma,\quad \Pi^\mu_{\;\nu}\equiv \delta^\mu_{\;\nu}+\ell^\mu n_\nu,
\ee
where $n_\mu$ is the null normal to the horizon and $\ell_\mu$ ia an auxiliary null vector  whose normalization is conveniently chosen to
satisfy $\ell_\mu n^\mu=-1$. Next, we define the rescaled traceless horizon curvature
\be
\theta^i_{\; j}\equiv \sqrt{\frac{\gamma}{\kappa^2}}\Sigma^i_{\; j},
\ee
where $\sqrt{\gamma}$ is the horizon area element, $\Sigma^i_{\; j}\equiv \Theta^i_{\; j}-\frac{1}{D-2}\Theta^l_{\; l} \delta^i_{\; j}$ is the traceless part of the extrinsic curvature and $\kappa$ is defined by the geodesic equation
 $n^\mu \nabla_\mu n_\nu=\kappa n_\nu$.
Then the horizon curvature power spectrum is defined as
\be\label{horizoncurpow}
\mathcal{A}(v,k)\equiv\frac{\partial}{\partial k}\int_{|\vec{k'}|\leq k} \frac{d^q k'}{(2\pi)^q}\tilde{\theta}^{*i}_{\;\; j}(v,\vec{k'})\,
\tilde{\theta}^{j}_{\; i}(v,\vec{k'}),
\ee
 with $\tilde{\theta}^i_{\; j}\equiv \int d^q x \,\theta^i_{\; j} e^{-i \vec{k}\cdot \vec{x}}$.

From (\ref{leadingordermetric}) and (\ref{horizonposition}), we obtain
\be
n_\mu dx^\mu=dr-\frac{1}{n}\Big(\frac{\partial_v m}{m} dv+\frac{\partial_a m}{m} dz^a \Big),\qquad \ell_\mu dx^\mu=-dt,
\ee
and
\be
\sqrt{\gamma}=\frac{m(v,z^a)}{n^{n/2}},\qquad \kappa=\frac{n}{2},
\ee
from which we have
\be
\theta^i_{\;j}=\frac{m \delta^{il}}{n^{n/2}}\Bigg[\frac{2\beta}{\beta+1}\Bigg(\partial_l\Big(\frac{p_j}{m}\Big)+\partial_j\Big(\frac{p_l}{m}\Big)\Bigg)
-\Bigg(\partial_l\Big(\frac{\partial_j m}{m}\Big)+\partial_j\Big(\frac{\partial_l m}{m}\Big)\Bigg)\Bigg].
\ee
Let us take the incompressible limit in which $m$ is treated as a constant, then
\be
\theta^i_{\;j}\simeq\frac{1}{n^{n/2}}\frac{2\beta}{\beta+1}\Big(\partial^ip_j+\partial_j p^i\Big),
\ee
and
\be
\tilde{\theta}^i_{\; j}\simeq\frac{1}{n^{n/2}}\frac{2\beta}{\beta+1}\Big(i k^i \tilde{p}_j+i k_j \tilde{p}^i\Big),
\ee
where $\tilde{p}_i$ is the Fourier transformation of $p_i$. Plugging this into (\ref{horizoncurpow}) and comparing with (\ref{energyspectrum}) we find
\be
\frac{\mathcal{A}(v,k)}{E_C(v,k)}\simeq\frac{2}{n^{n}} \Big(\frac{2\beta}{\beta+1}\Big)^2 k^2.
\ee
Thus in the low Mach number case the emergent $k^2$ relation  holds  for EGB gravity as well, with the coefficient embodying  the effect of the GB term.
As in the case of the Einstein gravity\cite{Rozali1707}, if the Mach number is not small the simple relation  $\mathcal{A}\sim k^2 E_C$ is expected to be broken.

\section{Summary}\label{section:summary}

In this paper we studied the holographic hydrodynamics of the fluid flow in the Einstein-Gauss-Bonnet gravity by using the large $D$ effective theory.
After performing the $1/D$ expansion and integrating out the radial direction of the EGB equations, we obtained the effective equations for the EGB-AdS black branes.
We found that the effective equations can be easily turned into the form of the equations for a dynamical fluid, the properties of which are  consistent with
the results obtained from the other studies in AdS/CFT. 

In the small Mach number limit, we found that the holographic hydrodynamic  fluid equations in the EGB gravity could reduce to the Navier-Stokes equations for the incompressible fluid, with the modified  Reynolds number and the modified Mach number
\be\label{redfReM}
\bRe=\frac{\beta+1}{\beta^{2}+1} L_0\, U,\quad \bM=\sqrt{\frac{\beta+1}{2}}U.
 \ee
Therefore, as in the case of the Einstein gravity, the Kolmogorov scaling laws are expected to emerge for three dimensional fluid flow, and an inverse/direct  cascade should  exhibit for two dimensional turbulent fluid flow.

 Compared to the hydrodynamical equations derived in the Einstein gravity, there are two extra terms in the equations of motion (\ref{Eq:dimenless2}), being proportional to $1/\bRe$. In three dimensions, the vortex stretching term has the dominant influence on the kinetic energy transfer between different scales, and it leads to the Kolmogorov scaling. In this case, the two extra terms play minor role. In two dimensions, as the vortex stretching term is vanishing, and the evolution of the enstrophy  is  determined by the palinstrophy-like terms, which include the contribution from the extra terms. One may expect that the 2D turbulence may present different feature due to the extra terms.

 Surprisingly, we found quite similar behaviors for the 2D holographic turbulence in the EGB theory as the one in the Einstein gravity.  We generated the turbulent flow in a toroidal domain by numerically solving the equations of motion. The evolution of the flow can be divided into three stages: the initial growth of the instabilities, the turbulent regime and the late time decay. Qualitatively,
 in all these stages the behavior of the EGB fluid flow is similar to the one in  the Einstein gravity: the initial instability grows accompanied with the formation of filaments, then the flow enters the turbulence regime in which the large scale structure is formed, and finally the formed counter-rotating vortices decay slowly. The formation of the large scale structure indicates an inverse cascade and the formation of filaments indicates the direct cascade. We found that the energy spectrum of the direct cascade obeys a $k^{-4}$ power law but the $k^{-5/3}$ law for the inverse cascade is absent. A possible reason for this could be the insufficient numerical precision \cite{Mininni2013}.

 Despite the qualitative similarity, the EGB turbulence presents some different features. The effect of  the GB term is twofold. On the one hand, the GB term affects the initial Reynolds number and Mach number. If the initial conditions are fixed,
  for a positive GB coefficient, both the Reynolds number and Mach number are smaller  than those of the fluid in the Einstein gravity,  such that the EGB fluid needs more time to reach the turbulent phase. However, for a negative GB coefficient, the Reynolds number is smaller but the Mach number is larger,
  so that the fluid needs less time to reach the turbulent phase.  On the other hand,   if we keep the initial Reynolds number and Mach number fixed,  the extra terms in the dynamical equation affect the dissipation rate and so the evolution rate of the fluid flow.
  Moreover, we found that the critical Reynolds number is not sensitive to the value of the Gauss-Bonnet coefficient.

Due to the simplification  at large $D$, we  showed analytically that in the regime of small Mach number the proposed relation between the horizon curvature power spectrum and the hydrodynamic energy power spectrum  $\mathcal{A}/E_C\propto k^2 $ \cite{Adams1307}  still holds in the EGB gravity.

In this work, we only did spectrum analysis for the 2D holographic EGB turbulence. It would be interesting to consider the 3D case.  Even though a Kolmogorov cascade is expected in 3D case, it would be nice to show it explicitly.


\section*{Acknowledgments}

BC would like to thank the participants in ``Gravity - New perspectives from strings and higher dimensions" at Centro de Ciencias de Benasque Pedro Pascual for stimulated discussions. In particular, he thanks R. Emparan, V. Hubeny, M. Rangamani and A. Yarom for the discussion  which initiated this project. We are grateful to M. Rozali, E. Sabat and A. Yarom for the correspondence and the clarification on the numerical simulation. We thank Shan-Quan Lan for the discussion on the energy spectrum analysis.
B. Chen and P.-C. Li
were supported in part by NSFC Grant No. 11275010, No. 11325522, No. 11335012
and No. 11735001. Y. Tian is partly supported by the National Natural Science Foundation of China (Grant Nos. 11475179 and 11675015) and also supported by the ``Strategic Priority Research Program of the Chinese Academy of Sciences'', grant No. XDB23030000. C.-Y. Zhang is supported by National Postdoctoral
Program for Innovative Talents BX201600005.


\begin{thebibliography}{}
\bibitem{Emparan1302} R. Emparan, R. Suzuki, and K. Tanabe, ``The large $D$ limit of General Relativity," JHEP {\bf1306} (2013) 009 [arXiv:1302.6382].

\bibitem{Emparan1402} R. Emparan, R. Suzuki and K. Tanabe, ``Instability of rotating black holes: large $D$ analysis," JHEP {\bf06}
(2014) 106, [arXiv:1402.6215].

\bibitem{Emparan1406} R. Emparan, R. Suzuki and K. Tanabe, `` Decoupling and non-decoupling dynamics of large $D$ black holes," JHEP \textbf{1407}, 113 (2014) [arXiv:1406.1258].

\bibitem{Emparan1502} R. Emparan, R. Suzuki and K. Tanabe, ``Quasinormal modes of (Anti-)de Sitter black holes in the $1/D$
expansion," JHEP {\bf04} (2015) 085, [arXiv:1502.02820].


\bibitem{Emparan1504} R. Emparan, T. Shiromizu, R. Suzuki, K. Tanabe, and T. Tanaka, `` Effective theory of Black Holes in the $1/D$ expansion," JHEP \textbf{06} (2015) 159, [arXiv:1504.06489].

\bibitem{Suzuki1505} R. Suzuki and K. Tanabe, ``Stationary black holes: Large $D$ analysis," JHEP \textbf{1509}, 193(2015) [arXiv:1505.01282].

\bibitem{Suzuki1506}R. Suzuki and K. Tanabe, `` Non-uniform black strings and the critical dimension in the
$1/D$ expansion," JHEP {\bf10} (2015) 107, [arXiv:1506.01890].

\bibitem{Emparan1506}  R.~Emparan, R.~Suzuki and K.~Tanabe, ``Evolution and End Point of the Black String Instability: Large $D$ Solution,"
  Phys.\ Rev.\ Lett.\  {\bf 115}, no. 9, 091102 (2015) [arXiv:1506.06772].

\bibitem{Tanabe1510}  K.~Tanabe,  ``Black rings at large $D$,"  JHEP {\bf 1602}, 151 (2016)  [arXiv:1510.02200].

\bibitem{Chen1702} B. Chen, P.-C. Li and Z. z. Wang, ``Charged Black Rings at large D," JHEP {\bf1704} (2017) 167,
[arXiv:1702.00886].

\bibitem{Tanabe1511} K. Tanabe, ``Instability of de Sitter Reissner-Nordstrom black hole in the $1/D$ expansion," Class. Quant. Grav. \textbf{33} no. 12, 125016 (2016) [arXiv:1511.06059].

\bibitem{Rozali1607} M. Rozali and A. Vincart-Emard, ``On Brane Instabilities in the Large $D$ Limit," doi:10.1007/JHEP08(2016)166 [arXiv:1607.01747].

\bibitem{Miyamoto1705} U. Miyamoto, ``Non-linear perturbation of black branes at large $D$," JHEP {\bf06} (2017) 033, [arXiv:1705.00486].

\bibitem{Emparan:2018bmi}
  R.~Emparan, R.~Luna, M.~Martinez, R.~Suzuki and K.~Tanabe,
  ``Phases and Stability of Non-Uniform Black Strings,''
  arXiv:1802.08191 [hep-th].

\bibitem{Bhattacharyya1504} S. Bhattacharyya, A. De, S. Minwalla, R. Mohan and A. Saha, `` A membrane paradigm at large $D$," JHEP \textbf{1604}, 076 (2016) [arXiv:1504.06613].

\bibitem{Bhattacharyya1511}S. Bhattacharyya, M. Mandlik, S. Minwalla and S. Thakur, ``A Charged Membrane Paradigm at Large $D$," JHEP \textbf{1604}, 128 (2016) [arXiv:1511.03432].



\bibitem{Dandekar1607}  Y.~Dandekar, A.~De, S.~Mazumdar, S.~Minwalla and A.~Saha,  ``The large $D$ black hole Membrane Paradigm at first subleading order,''  JHEP {\bf 1612}, 113 (2016)[arXiv:1607.06475].

\bibitem{Dandekar1609}  Y.~Dandekar, S.~Mazumdar, S.~Minwalla and A.~Saha,  ``Unstable `black branes' from scaled membranes at large $D$,''  JHEP {\bf 1612}, 140 (2016) [arXiv:1609.02912].

\bibitem{Bhattacharyya1704} S. Bhattacharyya, P. Biswas, B. Chakrabarty, Y. Dandekar and A. Dinda, ``The large $D$ black hole
dynamics in AdS/dS backgrounds," [1704.06076].

\bibitem{Dandekar:2017aiv}
  Y.~Dandekar, S.~Kundu, S.~Mazumdar, S.~Minwalla, A.~Mishra and A.~Saha,
  ``An Action for and Hydrodynamics from the improved Large D membrane,''
  arXiv:1712.09400 [hep-th].

\bibitem{Damour:1978cg}
  T.~Damour,
  ``Black Hole Eddy Currents,''
  Phys.\ Rev.\ D {\bf 18}, 3598 (1978).

 \bibitem{Thorne:1986iy}
  K.~S.~Thorne, R.~H.~Price and D.~A.~Macdonald,
  ``Black Holes: The Membrane Paradigm,''
  NEW HAVEN, USA: YALE UNIV. PR. (1986) 367p.


\bibitem{Emparan1602}R. Emparan, K. Izumi, R. Luna, R. Suzuki and K. Tanabe, `` Hydro-elastic complementarity in black branes at large $D$," JHEP \textbf{06} (2016) 117, [arXiv:1602.05752].

\bibitem{Bhattacharyya0712} S. Bhattacharyya, V. E. Hubeny, S. Minwalla and M. Rangamani, ``Nonlinear Fluid Dynamics from
Gravity," JHEP {\bf02} (2008) 045, [arXiv:0712.2456].

\bibitem{Adams1307} A. Adams, P. M. Chesler, and H. Liu, ``Holographic turbulence," Phys. Rev. Lett. {\bf112}, 151602 (2014), [arXiv:1307.7267].

\bibitem{Carrasco1210} F. Carrasco, L. Lehner, R. C. Myers, O. Reula, and A. Singh, ``Turbulent flows for relativistic conformal fluids in 2+1 dimensions," Phys. Rev. {\bf D86}, 126006 (2012), [arXiv:1210.6702].

\bibitem{Green1309} S. R. Green, F. Carrasco and L. Lehner, ``A Holographic path to the turbulent side of gravity," Phys. Rev. X {\bf4}, 011001 (2014)
[arXiv:1309.7940].


\bibitem{Rozali1707} M. Rozali, E. Sabat and A. Yarom, ``Holographic Turbulence in a Large Number of Dimensions,"  [arXiv:1707.08973].

\bibitem{Zwiebach1985} B. Zwiebach, ``Curvature Squared Terms and String Theories," Phys. Lett. B \textbf{156} (1985) 315.

\bibitem{Boulware1985} D. G. Boulware and S. Deser, ``String-Generated Gravity Models," Phys. Rev. Lett. \textbf{55} (1985) 2656.


  \bibitem{Kovtun:2004de}
  P.~Kovtun, D.~T.~Son and A.~O.~Starinets,
  ``Viscosity in strongly interacting quantum field theories from black hole physics,''
  Phys.\ Rev.\ Lett.\  {\bf 94}, 111601 (2005)
  doi:10.1103/PhysRevLett.94.111601
  [hep-th/0405231].

\bibitem{Brigante:2007nu}
  M.~Brigante, H.~Liu, R.~C.~Myers, S.~Shenker and S.~Yaida,
  ``Viscosity Bound Violation in Higher Derivative Gravity,''
  Phys.\ Rev.\ D {\bf 77}, 126006 (2008)
  doi:10.1103/PhysRevD.77.126006
  [arXiv:0712.0805].

  \bibitem{Brigante:2008gz}
  M.~Brigante, H.~Liu, R.~C.~Myers, S.~Shenker and S.~Yaida,
  ``The Viscosity Bound and Causality Violation,''
  Phys.\ Rev.\ Lett.\  {\bf 100}, 191601 (2008)
  doi:10.1103/PhysRevLett.100.191601
  [arXiv:0802.3318].

  \bibitem{Hofman:2008ar}
  D.~M.~Hofman and J.~Maldacena,
  ``Conformal collider physics: Energy and charge correlations,''
  JHEP {\bf 0805}, 012 (2008)
  doi:10.1088/1126-6708/2008/05/012
  [arXiv:0803.1467].

\bibitem{Buchel0906} A. Buchel and R. C. Myers, ``Causality of Holographic Hydrodynamics," JHEP {\bf08} (2009) 016 [arXiv: 0906.2922].

\bibitem{Hofman:2009ug}
  D.~M.~Hofman,
  ``Higher Derivative Gravity, Causality and Positivity of Energy in a UV complete QFT,''
  Nucl.\ Phys.\ B {\bf 823}, 174 (2009)
  doi:10.1016/j.nuclphysb.2009.08.001
  [arXiv:0907.1625].

  \bibitem{Buchel0911} A. Buchel, J. Escobedo, R. C. Myers, M. F. Paulos, A. Sinha and M. Smolkin, ``Holographic GB gravity in arbitrary dimensions," JHEP {\bf{03}} (2010) 111[arXiv:0911.4257].

\bibitem{Chen1511} B. Chen, Z.-Y. Fan, P. Li and W. Ye, ``Quasinormal modes of Gauss-Bonnet black holes at large $D$," JHEP {\bf01} (2016) 085 [arXiv:1511.08706].

\bibitem{Chen1703} B. Chen and P.-C. Li, ``Static Gauss-Bonnet black holes at large $D$," JHEP {\bf1705} (2017) 025  [arXiv:1703.06381].

\bibitem{Chen1707} B. Chen, P.-C. Li and C.-Y. Zhang, ``Einstein-Gauss-Bonnet Black Strings at Large $D$," JHEP {\bf10} (2017) 123 [arXiv:1707.09766].

 \bibitem{Cai0109} R.-G. Cai,``Gauss-Bonnet Black Holes in AdS Spaces," Phys. Rev. D {\bf65}, 084014 (2002) [arXiv:hep-th/0109133].

 \bibitem{Davidson2015} P. Davidson, ``Turbulence: an introduction for scientists and engineers", Oxford University Press (2015).

\bibitem{Batchelor1969} G. K. Batchelor, ``Computation of the energy spectrum in homogeneous two-dimensional turbulence," The Physics of Fluids {\bf12} (1969) II-233.

\bibitem{Kraichnan1967}R. H. Kraichnan, ``Inertial ranges in two-dimensional turbulence," The Physics of Fluids {\bf10} (1967) 1417-1423.

\bibitem{Mininni2013} P. D. Mininni and A. Pouquet, ``Inverse cascade behavior in freely decaying two-dimensional fluid
turbulence," Phys. Rev. E {\bf87} (Mar, 2013) 033002.
\end{thebibliography}
\end{document}